\documentclass[nofootinbib,two
column,amsmath,amssymb,aps,prd,preprintnumbers,superscriptaddress]{revtex4-2}
\pdfoutput=1
\usepackage{graphicx}%
\usepackage{dcolumn}%
\usepackage{bm}%
\usepackage{multirow}
\usepackage{hyperref}
\usepackage[dvipsnames]{xcolor}
\usepackage{color}
\usepackage{pifont}
\usepackage{bbding}
\usepackage{dirtytalk}
\usepackage{bm}
\usepackage{subfig}
\usepackage{verbatim}
\usepackage{amsmath}
\usepackage{cases}
\usepackage{tensor}
\usepackage{appendix}
\usepackage{tabularx}

\newcommand{\til}{~}
\usepackage[T1]{fontenc} %
\usepackage{lipsum}  
\usepackage{tikz, pgfplots}
\usetikzlibrary{positioning}
\usetikzlibrary{fadings, shadings, shadows, positioning, shapes, calc}
\usepackage{ulem}
\usepackage{verbatim}
\usepackage{caption}
\usepackage{subcaption}
\usepackage{transparent}
\usepackage{adjustbox}

\tikzfading %
[
  name=fade out,
  inner color=transparent!0,
  outer color=transparent!100
]

\definecolor{grey}{rgb}{0.4,0.4,0.4}
\definecolor{dullmagenta}{rgb}{0.4,0,0.4}
\definecolor{darkblue}{rgb}{0,0,0.4}
\definecolor{midblue}{rgb}{0,0,0.5}
\definecolor{midred}{rgb}{0.5,0,0}
\definecolor{orange}{rgb}{1,0.5,0}
\definecolor{lightbrown}{rgb}{0.75,0.5,0.25}
\definecolor{tan}{cmyk}{0.14,0.42,0.56,0}
\definecolor{djunglegreen}{cmyk}{0.99,0,0.52,0}
\definecolor{lightgreen}{rgb}{0,1,0}
\definecolor{olivegreen}{cmyk}{0.64,0,0.95,0.40}
\definecolor{midgreen}{rgb}{0.0,0.675,0.0}
\definecolor{darkgreen}{rgb}{0,0.5,0}
\definecolor{ceruleanblue}{rgb}{0.0, 0.45, 1} 
\definecolor{burgundy}{rgb}{0.5, 0.0, 0.13}
\definecolor{hvred}{RGB}{186,12,47}
\definecolor{ste}{rgb}{0.01, 0.28, 1.0}

\hypersetup{
    colorlinks=true,
    linkcolor=ceruleanblue,
    filecolor=ceruleanblue,      
    urlcolor=ceruleanblue, %
    citecolor=midblue, %
}

\newcolumntype{Y}{>{\centering\arraybackslash}X}

\begin{document}
\title{Gravitational wave propagation beyond General Relativity: \\ geometric optic expansion and lens-induced dispersion} 

\author{Nicola Menadeo}
\email{nicola.menadeo@aei.mpg.de}
\affiliation{Scuola Superiore Meridionale, Largo San Marcellino 10, I-80138, Napoli, Italy}
\affiliation{INFN Sezione di Napoli, Compl. Univ. di
Monte S. Angelo, Edificio G, Via Cinthia, I-80126, Napoli, Italy}
\affiliation{Max Planck Institute for Gravitational Physics (Albert Einstein Institute) \\
Am Mühlenberg 1, D-14476 Potsdam-Golm, Germany}

\author{Miguel Zumalac\'arregui}
\email{miguel.zumalacarregui@aei.mpg.de}
\affiliation{Max Planck Institute for Gravitational Physics (Albert Einstein Institute) \\
Am Mühlenberg 1, D-14476 Potsdam-Golm, Germany}

\begin{abstract}
The nature of gravity can be tested by how gravitational waves (GWs) are emitted, detected, and propagate through the universe. Propagation tests are powerful, as small deviations compound over cosmological distances. 
However, GW propagation tests of theories beyond Einstein's general relativity (GR) are limited by the high degree of symmetry of the average cosmological spacetime. 
Deviations from homogeneity, \textit{i.e.} gravitational lenses, allow for new interactions, \textit{e.g.,} between standard GW polarization and new scalar or vector fields, with different spin. Therefore, \textit{GW lensing beyond GR} offers novel tests of cosmological gravity.
Here we present the theory of GW propagation beyond GR in the short-wave expansion, including corrections to the leading-order amplitude and phase for the first time. 
As an example, we compute the dispersive (frequency-dependent) corrections to all metric and scalar field perturbations in Brans-Dicke, the simplest modified theory exhibiting GW dispersion.
GW lensing effects are too small to observe in Brans-Dicke theories compatible with solar system and binary pulsar limits. Nevertheless, our formalism opens the possibility of novel tests of gravity, including dark-energy theories and screening mechanisms.
\end{abstract}

\maketitle
{\hypersetup{hidelinks}
  \tableofcontents}
\section{Introduction}

Einstein's General Relativity (GR) has been remarkably successful in describing gravitational phenomena on a wide range of systems, across vastly different scales. However, open questions remain on both extremely small and large scales. Small-scale/high-energy problems of GR include the nature of spacetime singularities and the quantum completion of the theory. On cosmological scales, GR has led to the need to include dark matter and dark energy, accounting for 95\% of the universe's content today~\cite{Planck:2018vyg,Oks:2021hef,PhysRevD.33.3495,2017AdSpR..60..166A}.
In particular dark energy, responsible for cosmic acceleration~\cite{SupernovaSearchTeam:1998fmf,SupernovaCosmologyProject:1998vns,Li:2012dt,Li:2011sd,SupernovaSearchTeam:1998fmf,Frieman:2008sn,2013PhR...530...87W,Copeland:2006wr}, could be interpreted as a breakdown of the attractive nature of gravity on the largest scales. This hypothesis has resulted in a widespread investigation of alternative theories~\cite{Clifton:2011jh,Capozziello:2011et,Joyce:2014kja}, as well as an ambitious observational program to test them using cosmological observations~\cite{Weinberg:2013agg,Planck:2015bue,Euclid:2024yrr,Cusin:2017mzw,Cusin:2017wjg,Bellini:2014fua,Zumalacarregui:2016pph,Ishak:2018his,Abdalla:2022yfr}.

The detection of GWs~\cite{LIGOScientific:2016aoc} opened a new frontier in the study of gravitational physics, both illuminating the foundations of GR and casting away alternative theories to the shadows. Observed GW signals are emitted by relativistic compact objects, thus probing the regime of strong-field and dynamical gravity~\cite{LIGOScientific:2019fpa,LIGOScientific:2020tif,LIGOScientific:2021sio}. In addition to emission, GW propagation across the universe are highly sensitive to other properties of gravity, such as the GW speed~\cite{LIGOScientific:2017zic} and the graviton mass~\cite{deRham:2016nuf}. GW propagation tests probe gravitational interactions directly and can achieve high sensitivity, as anomalous effects accumulate over cosmological distances. 
Most crucially, they apply directly to theories that modify cosmological dynamics, constraining many dark-energy theories~\cite{Ezquiaga:2017ekz,Creminelli:2017sry,Baker:2017hug,Lombriser:2015sxa,Bettoni:2016mij,Bamba:2012cp,Sakstein:2017xjx,Creminelli:2018xsv,Creminelli:2019kjy,Creminelli:2019nok,Lagos:2024boe,Saltas:2014dha}. Many GW propagation tests, including measuring the anomalous speed and amplitude, are limited by require an electromagnetic counterpart or otherwise determining the redshift of the source. Moreover, only a handful of effects exist on the homogeneous and isotropic Friedman-Robertson-Walker (FRW) cosmological background~\cite{Ezquiaga:2018btd,Ezquiaga:2021ler} that describes the average universe.

GW propagation over inhomogeneous spacetimes, $i.e.$ GW lensing, drastically extend the range of phenomena that can be used to test gravity. 
The basic principle is that lenses/inhomogeneities break the FRW symmetries, allowing interactions between fields of different spin. In alternative theories, this means that the GR standard degrees of freedom (d.o.f.), the $+,\times$ metric polarizations, mix with new fields such as scalars or vectors.
In addition to the many gravitational lensing effects in GR~\cite{Bartelmann:2010fz,Takahashi:2005sxa,Schneider:1992bmb,Bozza:2001xd,Bonga:2024orc,Takahashi:2003ix,Tambalo:2022plm,Oancea:2020khc,Oancea:2022szu,Oancea:2023hgu,Kubota:2024zkv,Braga:2024pik,Kubota:2023dlz,Villarrubia-Rojo:2024xcj}, lens-induced effects provide a powerful discriminant between theories through novel wave-optic pheonomena.
Interactions with new fields involving two derivatives cause lens-induced birefringence (LIB), a difference in speed between the $+$ and $\times$ components of GWs~\cite{Ezquiaga:2020dao}. The absence of birefringence in GW data provided constraints on cosmological theories of gravity comparable to those based on the GW speed~\cite{Goyal:2023uvm}. Related birefringent effects are predicted in strong gravitational fields in GR~\cite{Andersson:2020gsj,Oancea:2022szu,Oancea:2023hgu}, as well as in parity-violating theories~\cite{Wang:2021gqm}.
Although progress has been made to characterize GW lensing beyond GR~\cite{Ezquiaga:2020dao,Dalang:2019rke,Dalang:2020eaj,Takeda:2024ghe,Streibert:2024cuf,Faraoni:1998sp}, computational difficulties and the wide landscape of theories have prevented a systematic characterization of these phenomena.

One of the ways in which the theory of GW lensing beyond GR needs to be developed is by including frequency-dependent propagation effects. 
Most analyses rely on the \textit{geometric optics} (GO) approximation \til\cite{Garoffolo:2019mna,Dalang:2020eaj,Dalang:2019rke,Kubota:2022lbn,Koksbang:2021alx,Andersson:2020gsj,Tasinato:2021wol,Hou:2019wdg,Faraoni:1996rc}, which holds when the signal's wavelength is significantly smaller than the local curvature scale, the $i.e.$ short-wave approximation. In this regime, GWs travel along null geodesics, and their polarization is parallel-transported along them\til\cite{Andersson:2020gsj,Koksbang:2021alx,Oancea:2020khc}. 
Corrections \textit{beyond geometric optics} (bGO)\til\cite{Cusin:2019rmt,Dalang:2021qhu,Harte:2018wni,Harte:2019tid,Braga:2024pik,Santana:2024tch} include lens-induced dispersion (LID) on GW signals. Such modifications, by the nature of the short-wave approximation, are frequency-dependent and can be probed by GW interferometers.%
The characterization of the bGO regime beyond GR remains an open problem. A significant step forward was the first explicit computation of bGO corrections in GR\til\cite{Cusin:2019rmt,Dalang:2021qhu}, showing how bGO effects modify the amplitude and phase of GWs and discussing the emergence of apparent additional scalar polarizations modes, absent in the geometric optics limit. GW lensing beyond GR offers potential for even richer phenomena due to the presence of additional degrees of freecom~\cite{Goyal:2020bkm}.

In this work, we extend the study of GW lensing beyond GR, developing a formalism that incorporates bGO corrections that describe dispersive phenomena. For simplicity, we focus on the Brans-Dicke theory as an example of scalar-tensor models~\cite{Brans:1961sx,Faraoni:1999yp}. Using the short-wave expansion approximation, we compute the bGO corrections to the leading-order scalar observables for tensor and scalar waves passing through a point-like gravitational lens.

The work is organized as follows:\\
In Sec.\til\ref{sec:general_theory} we will introduce a general framework describing the full-propagation of the gravitational and scalar radiation. By employing the short-wave approximation we will show the general equations, order by order in the expansion parameter, the equations governing the geometric optics regime and the first corrections to it. Sec.\til\ref{GR} will be devoted to review calculations for GR, thus introducing the null-tetrad formalism and thereby presenting the geometric and beyond geometric optics equations and the respective formal solutions for the gravitational radiation. In Sec.\til\ref{BD} we will present the full propagation in the BD theory and show the solution in the geometric optics regime of the leading-order scalar and tensor amplitude. Then we will present, for the first time up to our knowledge, the general expression describing beyond geometric optics (bGO) corrections. In Sec.\til\ref{Point like lens}, we will explicitly evaluate the dispersive bGO corrections, analytically, in the special case of a point-like lens, reviewing the GR case shown in Ref.\til\cite{Dalang:2021qhu} and subsequently extending to BD.

\textit{Notation.}  Table.\til\ref{Tab: notation} sumarizes the main definitions that we will use throughout the work. Note that we will employ two equivalent descriptions of the theory, the Jordan (JF) and Einstein frame (EF). We will work with $c=\hbar=1$. Symmetrized and anti-symmetrized indices will be denoted as $T_{(\mu\nu)}\equiv(T_{\mu\nu}+T_{\nu\mu})/2$ and $T_{[\mu\nu]}\equiv(T_{\mu\nu}-T_{\nu\mu})/2$.

\begin{table}[h!]
\centering
\begin{tabular}{|c|c|c|}
\hline
 &Fields & Amplitude decomposition \\
\hline
GR& $\tilde{h}_{\mu\nu}$, Eqs.\til\eqref{eq: trace-reversed metric} & $\tilde{h}^{(n)}_{\mu\nu}\equiv \tilde{\alpha}^{(n)}_{AB}\Theta^{AB}_{\mu\nu}$, Eq.\til\eqref{amplitude decomposition} \\
\hline
EF& $(\tilde{h}_{\mu\nu},\delta\tilde{\phi})$, Eqs.\til\eqref{BD diagonalization transformation} & $\tilde{h}^{(n)}_{\mu\nu}\equiv \tilde{\alpha}^{(n)}_{AB}\Theta^{AB}_{\mu\nu}$, Eq.\til\eqref{amplitude decomposition} \\
\hline
JF & $(h_{\mu\nu},\delta\phi)$ & ${h}^{(n)}_{\mu\nu}\equiv {\alpha}^{(n)}_{AB}\Theta^{AB}_{\mu\nu}$\\
\hline
\end{tabular}
\caption{Notation used in the work.}
\label{Tab: notation}
\end{table}

\section{General theory for GW propagation}\label{sec:general_theory}

In any theory of gravity, the propagation of GWs is determined by the equations of motion (EoM) for linearized perturbations, obtained by expanding around the background metric
\begin{equation}\label{metric perturbation}
    g_{\mu\nu}^{\text{tot}}=g_{\mu\nu}+h_{\mu\nu}.
\end{equation}
Scalar-tensor theories of gravity include an additional scalar field which can be similarly expanded around the background solution as 
\begin{equation}\label{scalar field perturbation}
    \phi^{\text{tot}}=\bar{\phi}+\delta\phi.
\end{equation}
The evolution of both gravitational and scalar waves can be then described by a set of coupled differential equations, which can be compactly written as
\begin{equation}
    \mathbf{D}_{IJ}V^J=0,
\end{equation}
with
\begin{equation}\label{eq:EOM}
    \mathbf{D}_{IJ}\equiv \mathcal{K}_{IJ}^{\alpha\beta}\nabla_{\alpha}\nabla_{\beta}+\mathcal{A}_{IJ}^{\alpha}\nabla_{\alpha}+\mathcal{M}_{IJ},
\end{equation}
where $\nabla_{\mu}$ identifies the covariant derivative compatible with the background metric tensor. $V^J\equiv(h_{\mu\nu},\delta\phi)$ is a vector whose components are the dynamical field perturbations of the metric and the scalar field, after imposing the constraints on the non-dynamical components of $h_{\mu\nu}$, $e.g.,$~\cite[Sec.~V-B]{Ezquiaga:2020dao}.
Here, Latin capital indices denote components of the field perturbation, while Greek indices denote spacetime components.

The propagation equation \eqref{eq:EOM} is naturally split into three parts:  the \textit{kinetic matrix}, $\mathcal{K}_{IJ}^{\alpha\beta}$ encodes the second-order differential operator that acts on field perturbations. 
 \textit{Amplitude matrix}, $\mathcal{A}_{IJ}^{\alpha}$, contains terms with first-order derivatives. 
Finally, the \textit{mass matrix}, $\mathcal{M}_{IJ}$, involves the contributions of zeroth-order derivative terms.
More precisely, the vector $V^J$ formally has $11$ components (the usual 10 components of the metric plus one of the scalar field), and the respective matrices will be $11\times11$.

The propagating d.o.f. are determined by diagonalizing the kinetic matrix, and are known as \textit{propagation eigenstates}. The diagonalization is position-dependent and can be quite involved in general.
For the scalar-tensor theories of interest in this work, the $i.e.$ BD theory, diagonalization can be performed covariantly\til\cite{Ezquiaga:2020dao,Dalang:2020eaj}. Moreover, the transformation that achieves the diagonalization is equivalent to the well-known mapping between the Jordan and Einstein frames \cite{Flanagan:2004bz,Zumalacarregui:2012us,Bettoni:2013diz,Zumalacarregui:2013pma,Bettoni:2015wta,Capozziello:1996xg,Faraoni:1999hp}.
Once the kinetic matrix has been diagonalized, the eigenvectors of the evolving system correspond to the propagating eigenstates. The eigenvalues provide the dispersion relations which determine the propagation speed for each propagating d.o.f.~.
Hence, one can formally write
\begin{equation}\label{Diagonalization Kinetic matrix general}
\mathcal{K}^{\alpha\beta}_{IJ}e^I=\mathcal{G}^{\alpha\beta}_{I}e^I,
\end{equation}
where $e^I$ as an eigenstate (or polarization vector) of the kinetic matrix satisfying $e^Ie^J=\mathbf{\delta}^{IJ}$ and $\mathcal{G}^{\alpha\beta}_{I}$ is the \textit{effective metric} of the \textit{I-th} propagating eigenstate.

\begin{table*}[t]\label{tab:order_derivatives}
\vspace{-10pt}
  \centering
 \def\imwidth{2.5cm}
\begin{tabular}{|c | c c c | c|}
\hline
\textbf{Modified} & \multicolumn{3}{c|}{Geometric optics} & Wave optics \\[0pt]
 \textbf{GW lensing}& $\mathcal{O}(f^2)$ & $\mathcal{O}(f^0)$ & $\mathcal{O}(f^{-2})$ & Arbitrary $f$\\\hline
Effect & speed & amplitude & phase & all \\[2pt] \hline
Observable & birefringence & oscillations & dispersion &diffraction \\[2pt]
 & \includegraphics[width=\imwidth]{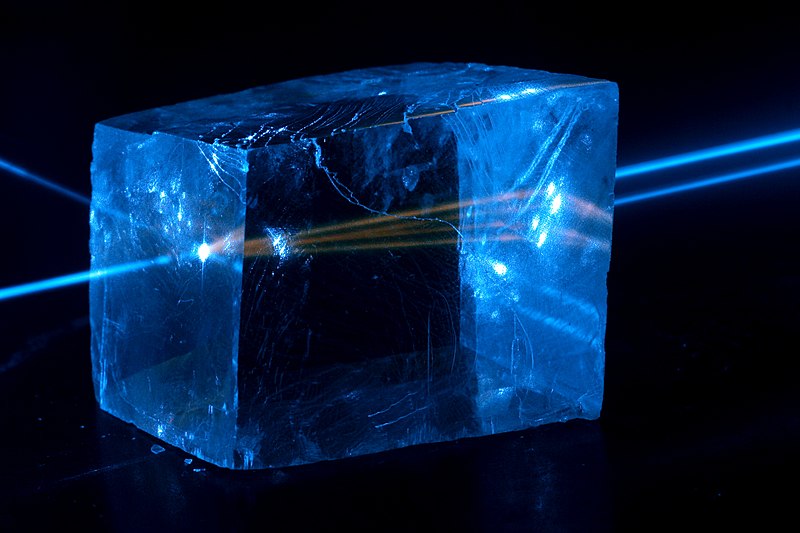} & \includegraphics[width=\imwidth]{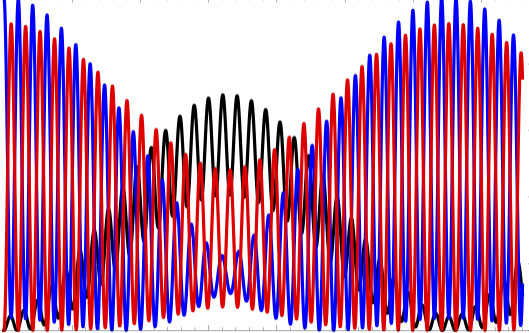} & \includegraphics[width=\imwidth]{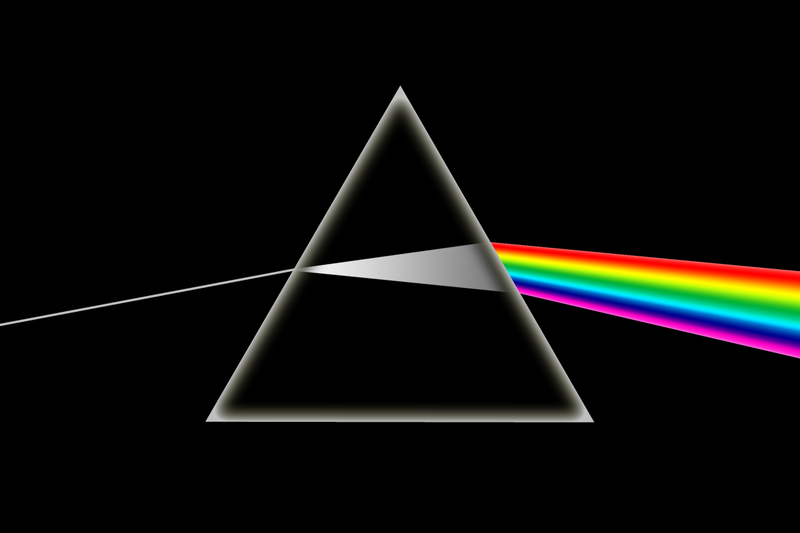} & \includegraphics[width=\imwidth]{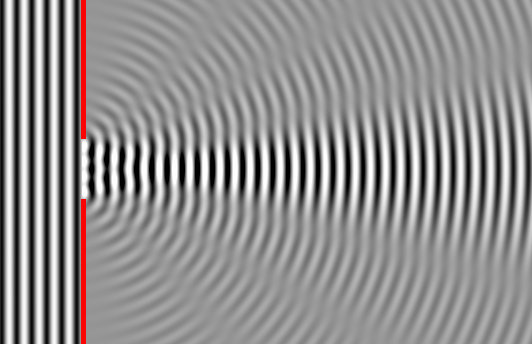} \\[2pt] \hline
Interactions Tested & 2 derivatives & 1 derivatives & 0 derivatives & all \\[2pt] \hline
Developed in  & Refs.~\cite{Ezquiaga:2020dao,Goyal:2023uvm} & \multicolumn{2}{|c|}{\textbf{this work} (for Brans-Dicke)} & future work\\ \hline
\end{tabular}
    \caption{Novel effects allowed by lens-induced interactions between $h_+,h_\times$ and additional fields. 
        ``Interactions tested'' refers to the number of derivatives in the couplings between GW polarizations ($h_+,h_\times$) and additional fields: lower (higher) numbers are more generic (restricted). %
    \label{tab:tgr}
     \vspace*{-0pt}
\vspace*{-15pt}
}
\end{table*}

\subsection{Short wave Expansion}\label{Short wave expasion general}

As a working hypothesis, we focus on the case in which the GW wavelength $\lambda$ is much smaller than the scale over which the spacetime varies significantly,  $\mathcal{R}_{\rm back}$.
In such conditions, we can perform the short-wave expansion (also known as WKB approximation) by expanding the perturbation fields as
\begin{equation}\label{WKB ansatz}
    V^I=e^{i\theta^{I}/\epsilon}\sum_n \epsilon^n A^{(n)I},
\end{equation}
where $\epsilon\equiv2\lambda/\mathcal{R}_{\rm back}$ is a dimensioneless book-keeping parameter and $\theta^{I}$ is the wave's phase related to the propagating d.o.f.~. The quantity $ A^{(n)I}$ contains information on both the amplitude and the polarization, and can be splitted as 
\begin{equation}\label{polarization}
    A^{(n)I}\equiv a^{(n)}e^I,
\end{equation}
where $a^{(n)}$ is a complex amplitude and $e^I$ the polarization vector. Moreover, the short-wave expansion naturally defines the scalar and tensor wavevectors, respectively
\begin{equation}\label{wavevector}
k^{I}_{\mu}\equiv\nabla_{\mu}\theta^{I}.
\end{equation}
By inserting the ansatz\til\eqref{WKB ansatz} into Eq.\til\eqref{eq:EOM} and retaining terms of equal order in $\epsilon$, we obtain the evolution equations for the phase and amplitude in the geometric optics regime, as well as the correction for these quantities in the bGO limit.

\subsubsection{Geometric optics}\label{General theory for gravitational radiation: geometric optics
}
The leading and next-to-leading order (NLO) equations in the short-wave expansion describe the geometric optic limit. The leading order equation is
\begin{equation}\label{eq: LO eq general framework}
\mathcal{G}_{I}^{\alpha\beta}k^{I}_{\alpha}k^{I}_{\beta}=0\,.
\end{equation}
It determines the evolution of the phase, and the speed of the propagation eigenstates, which in turn determine the theory's causal structure.

The diagonalization procedure\til\eqref{Diagonalization Kinetic matrix general} might lead to the case in which the resulting eigenvalues exhibit the same dispersion relation for each propagating eigenstate, thereby defining the theory as \textit{fully luminal}.%
\footnote{The term luminal commonly denotes the speed at which a tensor wave propagates, on FRW, is equal to the speed of light.} In this work we choose to focus on fully luminal theories.

The NLO contribution provides a differential equation for the evolution of the leading order amplitude
\begin{equation}\label{1st order correction}
\left[\mathcal{K}_{IJ}^{\alpha\beta}\left(2k^{J}_{\alpha}\nabla_{\beta}+\nabla_{\beta}k^{J}_{\alpha}\right)+\mathcal{A}_{IJ}^{\alpha}k^{J}_{\alpha}\right]A^{(0)J}=0.
\end{equation}

\subsubsection{Beyond geometric optics}\label{General Framework Beyond GO}
Let us now discuss the NLO corrections to the amplitude. This contribution is given by the equation at order $\epsilon^0$,
\begin{equation}\label{2nd order}
\left[\mathcal{K}_{IJ}^{\alpha\beta}\left(2k^{J}_{\alpha}\nabla_{\beta}+\nabla_{\beta}k^{J}_{\alpha}\right)+\mathcal{A}_{IJ}^{\alpha}k^{J}_{\alpha}\right]A^{(1)J}=i\mathbf{D}_{IJ}A^{(0)J}.
\end{equation}
The LHS of the above equation is identical to that of Eq.\til\eqref{1st order correction}. The RHS, instead, is given by the
full propagation equation acting on the leading order quantities. In other words, the failure of the leading order to satisfy the
full propagation equation sources the NLO amplitude. Higher-order corrections follow the same equation,
by simply substituting $A^{(1)J}$, $A^{(0)J}$ for $A^{(n)J}$, $A^{(n-1)J}$.

The leading bGO term describes a correction to the signal's phase. This interpretation follows from having an imaginary correction, $i A^{(1)J}$, to a real amplitude $A^{(0)J}$, $i.e.$ the imaginary unit in Eq.~\eqref{2nd order}.
This motivates rewriting Eq.\til\eqref{WKB ansatz} as
\begin{align}
V^J&=e^{i\theta^I/\epsilon}A^{(0)J}\left(1+i \epsilon \frac{A^{(1)J}}{A^{(0)J}}\right)\approx\\&\label{eq:explicit_frequency_phase_correction}\approx  e^{i\theta^I/\epsilon}A^{(0)J}e^{i\epsilon (A^{(1)J}/A^{(0)J})}
\\&
\equiv \label{eq:beta_phase}e^{i\theta^I/\epsilon}A^{(0)J}\exp\left(i\frac{\beta^{J}}{GMf}\right),
\end{align}
where the second line follows from $\mathcal{A}^{(1)J}\ll \mathcal{A}^{(0)J}$. The last line defines the dimensionless \textit{LID phasing parameter} $\beta^J$, where the signal frequency $f$ and lens' scale $M$ have been factored out~\cite{Oancea:2022szu,Oancea:2023hgu}.%

\begin{figure*}
  \begin{minipage}[h!]{\textwidth}
{\includegraphics[width=0.464\linewidth]{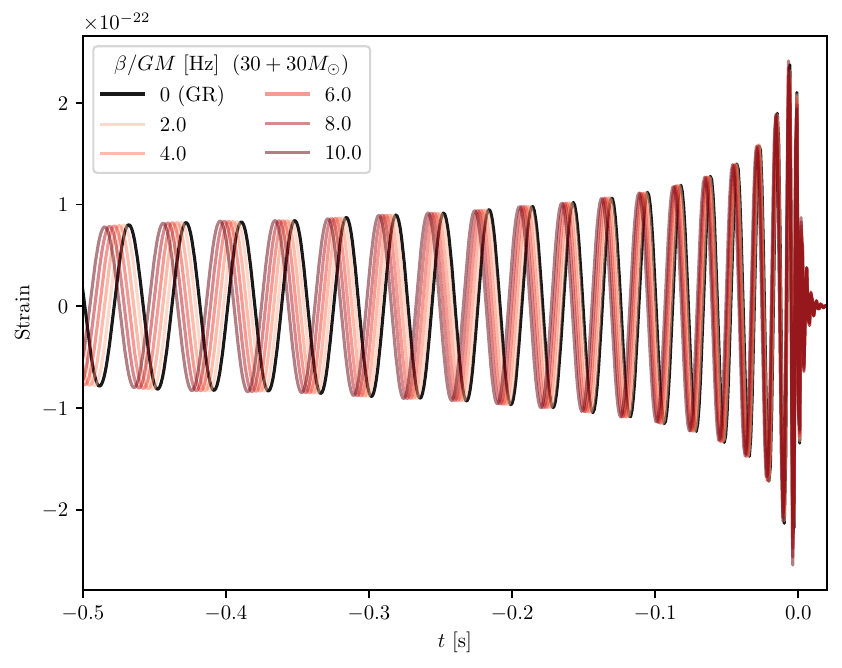}\label{fig:subfig2}}\hfill
{\includegraphics[width=0.464\linewidth]{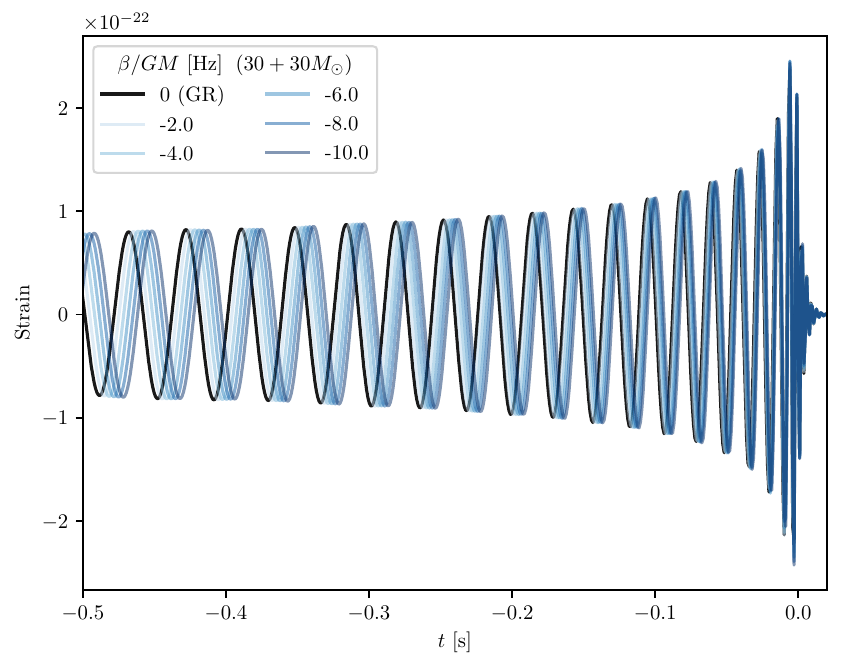}\label{fig:subfig1}}
  \end{minipage}
  \caption{Dispersive effects on GW signals. The plots show the LID corrections for positive (left) and negative (right) values of the phasing parameter $\beta$, and assuming that the effect is equal for both polarizations, $\beta^+=\beta^\times$. Deviations are most apparent in the inspiral phase, corresponding to lower frequencies. The signals correspond to a $30+30M_\odot$, non-spinning quasi-circular binary system.}
  \label{fig:waveforms}
\end{figure*}

LID produces a frequency- and polarization-dependent modulation of the waveform. Fig.~\ref{fig:waveforms} shows the effect on a typical source for positive/negative phasing, and assuming equal values for both polarizations, $\beta^+=\beta^\times$. Deviations are stronger in the inspiral phase, which corresponds to lower frequencies. 
The frequency-dependence of LID allows the effect to be tested on any GW signal, without the need of a electromagnetic counterpart.
On a given event, the correction is similar to a specific modified dispersion relation, $i.e.$ an apparent violation of Lorentz invariance. This deviation from GR is routinely tested on GW catalogs~\cite{LIGOScientific:2019fpa,LIGOScientific:2020tif,LIGOScientific:2021sio}, and for $\beta^J>0$ also equivalent to an effective graviton mass~\cite{Will:1997bb,deRham:2016nuf} (see Ref.~ \cite[Sec.~IV-A]{Oancea:2022szu} for details).

Two key differences allow one to distinguish LID from other deviations from GR. First, a violation of Lorentz invariance or graviton mass is universal: it will appear in all GW events, rather than on the fraction that is affected by a gravitational lens. Second, LID is polarization-dependent in general, $\beta^{I}\neq\beta^{J}$. 
LID effects in GR appear when GWs propagating in strong curvature regions, $i.e.$ a massive or intermediate-mass black hole acting as a lens~\cite{Oancea:2022szu,Oancea:2023hgu,Kubota:2024zkv}. LID due to strong-field GR effects can be distinguished by the detection of multiple images~\cite{Gondan:2021fpr,Leong:2024nnx} and the fact that the polarization dependence in GR (between the left-right polarized GW components) is suppressed by two additional powers of the frequency~\cite{Oancea:2022szu}.

In this work we will focus on gravity theories with a single additional scalar degree of freedom. However the framework can be generalized to any type of theory, regardless of the number and type of degrees of freedom.
In the following sections, we will specialize the calculations to GR and BD theory.

\subsubsection{Fully-luminal theories}
Note that Eqs.\til\eqref{1st order correction} and\til\eqref{2nd order} can be re-written by employing Eq.~\eqref{Diagonalization Kinetic matrix general}. Specifically, in certain scalar-tensor theories, the structure of $\mathcal{G}^{\alpha\beta}_I$ depends only on the background metric tensor; in this case, using Eq.~\eqref{eq: LO eq general framework} along with Eq.~\eqref{Diagonalization Kinetic matrix general}, the resulting eigenvalues yield the same dispersion relation for each propagating eigenstate, thus defining the theory as \textit{fully luminal}. In other words, such a definition states that scalar and tensor waves will propagate at the same speed. To this purpose, Eq.\til\eqref{eq: LO eq general framework} simplifies to 
\begin{equation}\label{eq:dispersion_rel_fully_luminal}
    g^{\mu\nu}k_\mu^I k_\nu^I=0, \hspace{0.2cm}\forall I.
\end{equation}
Further, Eqs.\til\eqref{1st order correction} and\til\eqref{2nd order}, respectively, take the following form
\begin{equation}
\left[2k^{J}_{\alpha}\nabla^{\alpha}+\nabla^{\alpha}k^{J}_{\alpha}+\mathcal{A}_{IJ}^{\alpha}k^{J}_{\alpha}\right]A^{(0)J}=0,
\end{equation}
\begin{equation}
\left[2k^{J}_{\alpha}\nabla^{\alpha}+\nabla^{\alpha}k^{J}_{\alpha}+\mathcal{A}_{IJ}^{\alpha}k^{J}_{\alpha}\right]A^{(1)J}=i\mathbf{D}_{IJ}A^{(0)J}.
\end{equation}
In this work, we choose to focus on fully luminal theories.

\section{General Relativity}\label{GR}
Let us start by describing the standard case of GR. We consider the perturbation on the Einstein tensor $G_{\mu\nu}\equiv R_{\mu\nu}-g_{\mu\nu}R/2$ providing the following linearized equation
\begin{equation}\label{eq: covariant propagation Eq GR}
\delta G_{\mu\nu}=\tensor{\mathsf{K}}{^{\alpha\beta}_{\mu\nu}}h_{\alpha\beta}+\mathsf{K}_{\mu\nu}h+\frac{1}{2}g_{\mu\nu}R^{\alpha\beta}h_{\alpha\beta}-\frac{1}{2}h_{\mu\nu}R=0,
\end{equation}
where $h\equiv g^{\alpha\beta}h_{\alpha\beta}$, $R_{\mu\nu}$ and $R$ are the trace, the Ricci tensor and scalar, respectively. Moreover,  $\tensor{\mathsf{K}}{^{\alpha\beta}_{\mu\nu}}$, $\tensor{\mathsf{K}}{_{\mu\nu}}$ are the component of the kinetic matrix%
\footnote{We will denote all terms contributing to the kinetic matrix as $\mathsf{K}$: different terms are uniquely specified by the number of spacetime indices, cf. Eqs.~\eqref{eq:K components GR1},~\eqref{eq:K components GR2}. A similar notation will be applied to $\mathsf{A}$, $\mathsf{M}$.} containing second order covariant derivative contribution
\begin{align}\label{eq:K components GR1}
    \tensor{\mathsf{K}}{^{\alpha\beta}_{\mu\nu}}&=-\frac{1}{2}\Box \delta^{\alpha}_{\mu}\delta^{\beta}_{\nu}+\nabla^{\alpha}\nabla_{(\mu}\delta^{\beta}_{\nu)}-\frac{1}{2}g_{\mu\nu}\nabla^{\alpha}\nabla^{\beta},\\ \label{eq:K components GR2}
    \mathsf{K}_{\mu\nu}&=\frac{1}{2}\Box-\frac{1}{2}\nabla_{\nu}\nabla_{\mu}.
\end{align}
Following the standard lore, the GWs propagation equation in vacuum is usually simplified by introducing the well-known trace-reversed metric perturbation and imposing the transverse and traceless (TT) gauge conditions on the latter~\cite{Maggiore:2007ulw,Misner:1973prb}, namely
\begin{align}\label{eq: trace-reversed metric}
    \tilde{h}_{\mu\nu}&\equiv h_{\mu\nu}-\frac{1}{2}g_{\mu\nu}h,\\ \label{transverse condition}
\nabla^{\mu}\tilde{h}_{\mu\nu}&=0,\\
\label{traceless condition}
\hspace{0.3cm} \tilde{h}&=0.
\end{align}
By plugging the above relations into Eq.\til\eqref{eq: covariant propagation Eq GR} and using the background Einstein field equation, one recovers the usual form of the wave equation
\begin{equation}\label{eq: Propagation GWs GR}
    \frac{1}{2}\Box \tilde{h}_{\mu\nu}-R_{\alpha\mu\nu\beta}\tilde{h}^{\alpha\beta}=0,
\end{equation}
with $R_{\alpha\mu\nu\beta}$ as the Riemann tensor.

Note that Eq.\til\eqref{eq: Propagation GWs GR} does not contain the trace $h$, which is removed by redefining the field as given by Eq.\til\eqref{eq: trace-reversed metric}. As we will see later, similar redefinitions will help simplify the analysis of GW propagation in theories beyond GR. %
The introduction of the trace-reversed perturbation metric, indeed, is essentially the field transformation leading to the kinetic matrix diagonalization discussed in Sec.\til\ref{Short wave expasion general}, thereby formally decoupling the evolution of the perturbation tensor field and the additional degree of freedom, specifically the trace in this case. Regardless of this, when working in vacuum (without the presence of matter), the trace can always be set to zero\til\cite{Misner:1973prb}.

\subsection{Short-wave expansion}

Let us now expand the field $\tilde{h}_{\mu\nu}$ using the WKB approximation as
\begin{equation}\label{WKB decomposition metric}
\tilde{h}_{\mu\nu}=\left(\tilde{h}^{(0)}_{\mu\nu}+\epsilon\tilde{h}^{(1)}_{\mu\nu}+...\right)e^{i \theta^T/\epsilon}\,.
\end{equation}
Here $\{\tilde{h}^{(0)}_{\mu\nu},\tilde{h}^{(1)}_{\mu\nu},...\}$ are the tensor amplitudes that can be further decomposed into a scalar coefficient and a polarization tensor, and $\theta^T$ is the phase related to the tensor wavevector $k_{\mu}^T$, according to Eq.\til\eqref{wavevector}. In the following we will denote $k_{\mu}^T\equiv k_{\mu}$ for simplicity.

\subsubsection{Geometric optics}

By plugging the ansatz\til\eqref{WKB decomposition metric} into Eq.\til\eqref{eq: Propagation GWs GR} one gets, at the leading order in $\epsilon$
\begin{equation}\label{Eq: dispersion relation GR}
    k^{\mu}k_{\mu}=0\,.
\end{equation}
This determines that $k^{\mu}$ is a null-vector and GWs propagate at the speed of light.

The NLO equation provides
\begin{equation}\label{eq: GR amplitude evolution}
\mathcal{D}\tilde{h}^{(0)}_{\mu\nu}=0,
\end{equation}\label{eq:transport_operator}
where
\begin{align}
\mathcal{D}\equiv\left(2k^{\alpha}\nabla_{\alpha}+\nabla_{\alpha}k^{\alpha}\right),
\end{align}
is the differential transport operator along the null ray, tangent to $k^{\alpha}$.
Further, plugging Eq.\til\eqref{WKB decomposition metric} into the transverse gauge condition\til\eqref{transverse condition} and keeping the leading-order terms, one obtains
\begin{equation}\label{eq: Transverse gauge WKB}
k^{\mu}\tilde{h}^{(0)}_{\mu\nu}=0\,.
\end{equation}
This indicates that the polarizations contained in $\tilde{h}^{(0)}_{\mu\nu}$ are orthogonal to the wavevector, and thus to the GWs propagation direction.

\subsubsection{Beyond geometric optics}

The equation describing the first order correction to the geometric optics is
\begin{equation}\label{eq: GR beyond geometric optics covariant}
     \mathcal{D}\tilde{h}^{(1)}_{\mu\nu}=i\left(\Box\tilde{h}^{(0)}_{\mu\nu}-2R_{\alpha\mu\nu\beta}\tilde{h}^{(0){\alpha\beta}}\right).
\end{equation}
Moreover, the transverse condition, at such an order, translates to
\begin{equation}\label{eq:transverse gauge bgo}
k^{\mu}\tilde{h}^{(1)}_{\mu\nu}=i\nabla^{\mu}\tilde{h}^{(0)}_{\mu\nu}.
\end{equation}
From the aforementioned equation it can be observed that bGO corrections are sourced from the leading-order amplitudes, which implies a deviation from orthogonality between polarizations, encoded in the bGO correction $h^{(1)}_{\mu\nu}$, and the propagation direction.

\subsection{Tetrad decomposition}\label{Tetrad decomposition}
In general, it is possible to decompose the amplitude in Eq.\til\eqref{WKB decomposition metric} by introducing a tetrad of null vectors adapted to $k^{\mu}$, where adapted means that every vector is parallely transported along the geodesic associated to $k^{\mu}$. The tetrad basis is given by
\begin{align}
        e^{\mu}_{A}&\equiv\{k^{\mu},m^{\mu},l^{\mu},n^{\mu}\},
\end{align}
where $n^{\mu}$ is real, $m^{\mu}$ and $l^{\mu}$ are complex such that
\begin{equation}\label{orthogonality null vector}
    l^{\mu}=\bar{m}^{\mu}, \hspace{0.3cm} g_{\mu\nu}m^{\mu}l^{\nu}=-g_{\mu\nu}k^{\mu}n^{\nu}=1,
\end{equation}
with all the other contraction vanishing.
The dual tetrad basis $\hat{e}^{\mu}_{A}$ can be defined as
\begin{equation}
    \hat{k}^{\mu}\equiv-n^{\mu},\hspace{0.2cm}\hat{n}^{\mu}\equiv-k^{\mu},\hspace{0.2cm}\hat{m}^{\mu}\equiv l^{\mu},\hspace{0.2cm}\hat{l}^{\mu}\equiv m^{\mu},
\end{equation}
such that $e^{\mu}_A\hat{e}^{B}_{\nu}=\delta^B_A\delta^{\mu}_{\nu}$.
Further, the background metric is decomposed as
\begin{equation}\label{eq:metric_tensor_decomposition}
    \bar{g}_{\mu\nu}=2m_{(\mu}l_{\nu)}-2n_{(\mu}k_{\nu)}.
\end{equation}

Generally, a rank-2 symmetric tensor can be decomposed along the null tetrad basis: in such a context, indeed, it is possible to split each tensor amplitude component (at any order $n$ of the short-wave expansion) as
\begin{equation}\label{amplitude decomposition}
    \tilde{h}^{(n)}_{\mu\nu}\equiv \tilde{\alpha}^{(n)}_{AB}\Theta^{AB}_{\mu\nu},
\end{equation}
where $\tilde{\alpha}^{(n)}_{AB}$ are the complex coefficients and $\Theta^{AB}_{\mu\nu}$ are the polarization tensors, constructed from the null tetrads and constituting the basis of the decomposition, defined as
\begin{equation}
    \Theta^{AB}_{\mu\nu}\equiv \frac{1}{2}\left(e^A_{\mu}e^B_{\nu}+e^A_{\nu}e^B_{\mu}\right).
\end{equation}

The expression in Eq.\til\eqref{amplitude decomposition} can be simplified by constraining some of the expansion coefficients using the TT gauge conditions.

\subsubsection{Geometric optics}

Specifically, by decomposing the leading order tensor amplitude $\tilde{h}^{(0)}_{\mu\nu}$ according to Eq.\til\eqref{amplitude decomposition} and plugging it into the leading order transversality condition\til\eqref{eq: Transverse gauge WKB}, one can readily verify that
\begin{equation}\label{eq:constrained leading amplitude}
\tilde{\alpha}^{(0)}_{nk}=\tilde{\alpha}^{(0)}_{nn}=\tilde{\alpha}^{(0)}_{nm}=\tilde{\alpha}^{(0)}_{nl}=0.
\end{equation}
The traceless condition\til\eqref{traceless condition}, further provides
\begin{equation}
    \tilde{\alpha}^{(0)}_{ml}=0.
\end{equation}
Still, we retain the freedom to fully fix the gauge. 
It is reasonable to expect that the spacetime is asymptotically flat, as the effect from the inhomogeneity is only localized around the lens and does not extend to infinity.
Consequently, one can transform the amplitude tensor to ensure the equation remains fully gauge invariant\til\cite{Cusin:2019rmt}, as follows
\begin{equation}
    \tilde{h}_{\mu\nu}\rightarrow \tilde{h}_{\mu\nu}+2C_{(\mu}k_{\nu)},
\end{equation}
where $C_{\mu}$ is an arbitrary complex vector orthogonal to $k_{\mu}$. 

Such a requirement is satisfied by imposing $n^{\mu}\tilde{h}^{(0)}_{\mu\nu}=0$, which, along with Eq.\til\eqref{amplitude decomposition}, provides
\begin{equation}
\tilde{\alpha}^{(0)}_{kk}=\tilde{\alpha}^{(0)}_{km}=\tilde{\alpha}^{(0)}_{kl}=0.
\end{equation}
After this process, the only unconstrained contributions are  $\tilde{\alpha}^{(0)}_{mm}$ and $\tilde{\alpha}^{(0)}_{ll}$, thus yielding
\begin{equation}\label{eq: Leading order tensor amplitude}
\tilde{h}^{(0)}_{\mu\nu}=\tilde{\alpha}^{(0)}_{mm}m_{\mu}m_{\nu}+\tilde{\alpha}^{(0)}_{ll}l_{\mu}l_{\nu}.
\end{equation}
These coefficients are the amplitudes associated with the two independent left-handed and right-handed helicity polarizations, typical for a massless spin-2 field, expressed in the null tetrad basis. 
The latter can be directly related to the usual GW amplitude polarizations $h_{+}$ and $h_{\times}$.

It is possible to express the above equation in terms of the usual tetrad basis~\til\cite{Isi:2022mbx,Will:2018bme}, upon which the standard GWs polarizations $h_+$ and $h_\times$ are defined, by performing a transformation on the tetrads 

\begin{align}
   \varepsilon^{+}_{\mu\nu}&\equiv \frac{ m_{\mu}m_{\nu}+l_{\mu}l_{\nu}}{\sqrt{2}},\\
   \varepsilon^{\times}_{\mu\nu}&\equiv i\frac{m_{\mu}m_{\nu}-l_{\mu}l_{\nu}}{\sqrt{2}},
\end{align}
where $\varepsilon^{\pm}_{\mu\nu}$ are the usual polarization tensors. Simultaneously, the coefficients change as follows 
\begin{align}\label{eq:hpluscross_alpha1}
    \tilde{\alpha}^{(0)}_{mm}&\equiv\frac{\tilde{h}^{(0)}_+ -i \tilde{h}^{(0)}_{\times}}{\sqrt{2}},\\ \label{eq:hpluscross_alpha2}
    \tilde{\alpha}^{(0)}_{ll}&\equiv\frac{\tilde{h}^{(0)}_+ +i \tilde{h}^{(0)}_{\times}}{\sqrt{2}}.
\end{align}

The evolution of $\tilde{\alpha}^{(0)}_{mm}$ and $\tilde{\alpha}^{(0)}_{ll}$ is obtained by inserting Eq.\til\eqref{eq: Leading order tensor amplitude} into Eq.\til\eqref{eq: GR amplitude evolution} and subsequently projecting along the dual tetrad basis $\hat{e}^{\mu}_{A}\hat{e}^{\nu}_{B}$, yielding a scalar equation
\begin{equation}\label{conserved amplitude GR}
\mathcal{D}\left(\tilde{\alpha}^{(0)}_{mm}\right)=\mathcal{D}\left(\tilde{\alpha}^{(0)}_{ll}\right)=0.
\end{equation}
By recalling $k^{\mu}\nabla_{\mu}\equiv d/d\xi$ as the directional derivative along $k^{\mu}$, with $\xi$ being an affine parameter, and using the relation $\nabla_{\mu}k^{\mu}=2d\ln D/d\xi$ where $D$ is the comoving distance along the geodesic (see Refs.\til\cite{Schneider:1992bmb,Poisson:2009pwt,Fleury:2015hgz} for a complete discussion on optical scalars), one can express Eq.\til\eqref{conserved amplitude GR} as a first order differential equation in the affine parameter
\begin{equation}
    \frac{d}{d\xi}\left[\tilde{\alpha}^{(0)}_{\circ}D\right]=0.
\end{equation}
The aforementioned equation can be easily integrated from the source point, labeled as $\xi_s$, to a generic point $\xi$ ($i.e.$ the observation point $\xi_o$) on the geodesic, thus providing
\begin{equation}
    \tilde{\alpha}^{(0)}_{\circ}(\xi)=\frac{D(\xi_s)}{D(\xi)} \tilde{\alpha}^{(0)}_{\circ}(\xi_s),
\end{equation}
where we can observe the standard evolution of the amplitudes decaying with increasing distance.
By plugging the above solution into Eq.\til\eqref{eq: Leading order tensor amplitude}, one can finally obtain the evolution, in the affine parameter, of the leading order tensor amplitude
\begin{equation}\label{Covariant Tensor amplitude GR final}
\tilde{h}^{(0)}_{\mu\nu}(\xi)=\frac{D(\xi_s)}{D(\xi)}\left(\tilde{\alpha}^{(0)}_{mm}(\xi_s)m_{\mu}m_{\nu}+\tilde{\alpha}^{(0)}_{ll}(\xi_s)l_{\mu}l_{\nu}\right).
\end{equation}

\subsubsection{Beyond geometric optics}

The evolution for the bGO coefficients $\tilde{\alpha}^{(1)}_{AB}$ of the tensor amplitude $h^{(1)}_{\mu\nu}$ (Eq.\til\eqref{amplitude decomposition}) can be computed with a similar approach.
An important difference is that some steps in the computation of
$\tilde{h}^{(0)}_{\mu\nu}$ do not apply to $h^{(1)}_{\mu\nu}$. Specifically, the transversality gauge condition in the bGO regime is determined by Eq.\til\eqref{eq:transverse gauge bgo}, which accounts for an orthogonality deviation between the polarizations of $h^{(1)}_{\mu\nu}$ and the propagation direction. By inserting the decomposition given by Eq.\til\eqref{amplitude decomposition} for $h^{(1)}_{\mu\nu}$ into Eq.\til\eqref{eq:transverse gauge bgo}, one gets
\begin{equation}\label{consistency equation}
2\tilde{\alpha}^{(1)}_{nn}n_{\mu}+\tilde{\alpha}^{(1)}_{nl}l_{\mu}+\tilde{\alpha}^{(1)}_{mn}m_{\mu}+\tilde{\alpha}^{(1)}_{nk}k_{\mu}=-2i \nabla^{\nu}\tilde{h}^{(0)}_{\mu\nu},
\end{equation}
which means that the above coefficients cannot be straightforwardly constrained to zero as it happens in Eq.~\eqref{eq:constrained leading amplitude} for $\tilde{h}^{(0)}_{\mu\nu}$. Moreover, the residual gauge condition $n^{\mu}\tilde{h}^{(1)}_{\mu\nu}=0$  and traceless one, $\tilde{h}=0$, provide 
\begin{equation}
     \tilde{\alpha}^{(1)}_{kk}=\tilde{\alpha}^{(1)}_{km}= \tilde{\alpha}^{(1)}_{kl}=\tilde{\alpha}^{(1)}_{kn}= \tilde{\alpha}^{(1)}_{ml}=0,
\end{equation}
ending up with
\begin{equation}\label{uncontrained tensor modes}
\begin{split}
\tilde{h}^{(1)}_{\mu\nu}&=\tilde{\alpha}^{(1)}_{mm}m_{\mu}m_{\nu}+\tilde{\alpha}^{(1)}_{ll}l_{\mu}l_{\nu}+\tilde{\alpha}^{(1)}_{nn}n_{\mu}n_{\nu}+\\&+\tilde{\alpha}^{(1)}_{nm}n_{(\mu}m_{\nu)}+\tilde{\alpha}^{(1)}_{nl}n_{(\mu}l_{\nu)}.
\end{split}
\end{equation}
The coefficients $\tilde{\alpha}^{(1)}_{nn}$, $\tilde{\alpha}^{(1)}_{nm}$, and $\tilde{\alpha}^{(1)}_{nl}$ are not constrained by the gauge conditions, unlike the leading-order amplitude in Eq.\til\eqref{eq:constrained leading amplitude}. 
This suggests these non-vanishing elements due to
bGO/dispersive effects appear as novel polarizations spreading the gravitational radiation. 

The evolution of the remaining $\tilde{\alpha}^{(1)}_{AB}$ coefficients is derived by inserting Eq.\til\eqref{amplitude decomposition} into Eq.\til\eqref{eq: GR beyond geometric optics covariant} and projecting along the dual tetrad basis $\hat{e}^{\mu}_{A}\hat{e}^{\nu}_{B}$, it is possible to derive and subsequently integrate the equation for the $\tilde{\alpha}^{(1)}_{AB}$ modes contributing to the leading order correction in the bGO regime, thus obtaining
\begin{equation}
    \begin{split}\label{BGO GR integration}
    \tilde{\alpha}_{AB}^{(1)}(\xi)&=\frac{D(\xi_s)}{D(\xi)} \tilde{\alpha}^{(1)}_{AB}(\xi_s)+\\&+\frac{i}{D(\xi)}\int_{\xi_s}^{\xi}d\xi \hat{e}^{\mu}_{A}\hat{e}^{\nu}_{B}D(\xi)\left[\Box\tilde{h}^{(0)}_{\mu\nu}-2R_{\alpha\mu\nu\beta}\tilde{h}^{(0){\alpha\beta}}\right].
\end{split}
\end{equation}
The first term describes the usual behavior of the mode propagating as $1/D$, while the second contribution encodes the dispersive, frequency-dependent corrections (as it will be explicitly evaluated later in Sec.\til\ref{Point like lens}). It is sourced by the background curvature, which can induce novel effects in the polarization tensor along the geodesic path. Further, the imaginary unit preceding the above integral indicates that the bGO corrections represent a modification of the GWs phase, rather than the (real-valued) leading-order amplitude\til(Eq.\til\eqref{eq:explicit_frequency_phase_correction}). This causes GW dispersive phenomena, $i.e.$ \textit{spectrum frequency decomposition} analog to the phenomenon observed when a light ray passes through an optical prism.

\section{Brans-Dicke}\label{BD}
The BD theory of gravity is arguably the simplest among scalar-tensor theories, featuring a scalar field non-minimally coupled to the Ricci scalar.
Assuming no interactions with any other matter fields, the theory is described by the action 
\begin{equation}
    S=\frac{1}{16 \pi G}\int d^4 x\sqrt{-g}\left(\phi R-\frac{\omega}{\phi}\phi_{\mu}\phi^{\mu}\right)+\int d^4 x\sqrt{-g}\mathcal{L}_m,
\end{equation}
where $g$ is the determinant of the metric tensor $g_{\mu\nu}$, $\mathcal{L}_m$ the matter Lagrangian, $\phi$ the scalar field coupled to gravity and $\omega$ the Brans-Dicke constant. Further,  ${\phi}_{\mu}\equiv \nabla_{\mu}{\phi}$. %
The field equations are easily obtained by varying the action w.r.t. the metric tensor and the scalar field, thus yielding
    \begin{align}\label{Eq: Tensor Field Eq.}
    \begin{split}
           \phi\left(R_{\mu\nu}-\frac{1}{2}g_{\mu\nu}R\right)&-\frac{\omega}{\phi}\phi_{\mu}\phi_{\nu}-\phi_{\mu\nu}+\\&+g_{\mu\nu}\left(\Box\phi+\frac{\omega}{2}\phi_{\alpha}\phi^{\alpha}\right)=8\pi GT_{\mu\nu},
    \end{split}
    \end{align}
    \begin{align}\label{Eq: Scalar Field Eq.}
        R+\frac{2\omega}{\phi}\Box\phi-\frac{\omega}{\phi^2}\phi_{\alpha}\phi^{\alpha}=0,
    \end{align}
where 
\begin{equation}
    T_{\mu\nu}\equiv\frac{-2}{\sqrt{-g}}\frac{\delta(\sqrt{-g}\mathcal{L}_m)}{\delta g^{\mu\nu}},
\end{equation}
is the energy-momentum tensor and $T\equiv g^{\mu\nu}T_{\mu\nu}$ its trace\til\cite{Bettoni:2015wta}.
By plugging Eq.\til\eqref{Eq: Tensor Field Eq.} and its contraction into Eq.\til\eqref{Eq: Scalar Field Eq.}, the scalar field equation simply reduces to
\begin{equation}
\Box\phi=\frac{8\pi G}{3+2\omega} T.
\end{equation}
In this work we will specialize to vacuum propagation $T_{\mu\nu}=0$.

The study of GWs propagation can be performed by perturbing the background metric and scalar field, as shown in Eqs.\til\eqref{metric perturbation} and\til\eqref{scalar field perturbation} and linearizing the above field equations, thus finding a coupled system of second-order differential equations governing the dynamics of the fields $(h_{\mu\nu},\delta\phi)$. In such a context, the diagonalization process of the kinetic matrix is feasible and can be performed covariantly by re-defining the metric perturbation as
\begin{align}\label{BD diagonalization transformation}
     \tilde{h}_{\alpha\beta}&\equiv h_{\mu\nu}-\frac{1}{2}g_{\mu\nu}h-g_{\mu\nu}\frac{\delta\phi}{\bar{\phi}},\\\delta\tilde{\phi}&=\delta\phi.
\end{align}
This redefinition is the linearized version of Einstein-frame metric, in which the scalar couples to matter, rather than curvature.

Moreover, having thus mapped the theory to the Einstein frame with this transformation, 
the TT gauge conditions, on $\tilde{h}_{\mu\nu}$, hold (See Ref.\til\cite{Lobato:2024rkb}).
The diagonalized propagation equation of perturbation fields in the Brans-Dicke theory, has the following structure
\begin{widetext}
\begin{equation}\til\label{eq: full progation Brans-Dicke}
\left[    \begin{pmatrix}
  \tensor{\mathsf{\Tilde{K}}}{_\mu_\nu^\alpha^\beta^\gamma^\rho} & 0 \\
  0 & \mathsf{\tilde{K}}^{\gamma\rho} \\
\end{pmatrix}\nabla_{\gamma}\nabla_{\rho}+
\begin{pmatrix}
 \tensor{\mathsf{\Tilde{A}}}{_{\mu\nu}^{\alpha\beta\gamma}} & \tensor{\mathsf{\Tilde{A}}}{_{\mu\nu}^{\gamma}} \\
  0 & \mathsf{\Tilde{A}}^{\gamma}\\
\end{pmatrix}\nabla_{\gamma}+
\begin{pmatrix}
 \tensor{\mathsf{\Tilde{M}}}{_{\mu\nu}^{\alpha\beta}} & \mathsf{\Tilde{M}}^{h}_{\mu\nu} \\
  \mathsf{\Tilde{M}}^{\alpha\beta}_{\phi} & \mathsf{\Tilde{M}}
 \\
\end{pmatrix}
\right]
\begin{pmatrix}
  \tilde{h}_{\alpha\beta} \\
  \delta\phi\\
\end{pmatrix}=0\,.
\end{equation}
\end{widetext}
We present the full expressions for the different $\mathsf{K,A, M}$ tensors in Appendix\til\ref{Terms Brans Dicke}.
\subsection{Short-wave expansion}\label{subsec:WKB-BD}
Similarly to the short-wave metric expansion, Eq.\til\eqref{WKB decomposition metric}, the perturbation of the scalar field can be also expanded as
\begin{align}
\label{WKB decomposition scalar}
\delta\phi\equiv\left(\delta\phi^{(0)}+\epsilon\delta\phi^{(1)}+...\right )e^{i\theta^S/\epsilon},
\end{align}
where $\{\delta\phi^{(0)},\delta\phi^{(1)},...\}$ are a set of scalar amplitudes and $\theta^S$ is the scalar phase. One can define $k^{S}_{\mu}$ as the wavevector related to $\theta^S$ following  Eq.\til\eqref{wavevector}.
Recalling the discussion in Sec.\til\ref{Short wave expasion general}, BD theory is classified as a fully luminal theory, which means that the scalar and tensor sectors share the same causal structure; once the gauge is fixed, the components of the diagonalized kinetic matrix, $i.e.$ $\tensor{\mathsf{\Tilde{K}}}{_\mu_\nu^\alpha^\beta^\gamma^\rho}$ and $\mathsf{\tilde{K}}^{\gamma\rho}$, are built only using the metric tensor. This leads to the same dispersion relation for scalar and tensor waves, resulting in the same propagation speed for all propagating eigenstates.
As a consequence, the scalar and tensor waves will share the same phase, and therefore we can set $\theta^{T}=\theta^{S}=\theta$, implying $k_{\mu}^{T}=k_{\mu}^{S}=k_{\mu}$.

In the following paragraphs, by inserting Eqs.\til\eqref{WKB decomposition metric} and\til\eqref{WKB decomposition scalar} into Eq.\til\eqref{eq: full progation Brans-Dicke}, we will analyze in detail the equations for both the scalar and tensor sectors, order by order in $\epsilon$.

\subsubsection{Geometric optics}

Let us start by studying the scalar wave propagation equation at the leading and next-to-leading order in $\epsilon$.
The relation at leading order is
\begin{equation}\label{dispersion relation scalar}
\mathsf{\tilde{K}}^{\alpha\beta}k_{\alpha}k_{\beta}=k_{\alpha}k^{\alpha}=0,
\end{equation}
where in the second equality we used\til\eqref{eq: BD effective metric}. As discussed in Sec\til\ref{GR}, Eq.\til\eqref{dispersion relation scalar} ensures that $k^{\mu}$ is a null vector propagating at the speed of light.

The NLO contribution describes the evolution of the leading order scalar amplitude and provides a differential equation for $\delta\phi^{(0)}$
\begin{equation}\label{LO scalar amplitude}
\left[\mathsf{\tilde{K}}^{\alpha\beta}\left(2k_{\alpha}\nabla_{\beta}+\nabla_{\alpha}k_{\beta}\right)-k_{\alpha}\mathsf{\Tilde{A}}^{\alpha}\right]\delta\phi^{(0)}=0.
\end{equation}
Using Eqs.\til\eqref{eq: BD effective metric} and\til\eqref{A gamma}, the above equation can be re-written in a more compact form
\begin{equation}\label{LO conservation Eq}
\mathcal{D}\left(\frac{\delta\phi^{(0)}}{\sqrt{\bar{\phi}}}\right)=0.
\end{equation}
Following the same procedure shown in Sec.\til\ref{Tetrad decomposition}, the latter can be expressed as
\begin{equation}
    \frac{d}{d\xi}\left(\frac{\delta\phi^{(0)}D}{\sqrt{\bar{\phi}}}\right)=0,
\end{equation}
which, in turn, can be easily integrated from the source of the wave $\xi_s$ to a generic point $\xi$, thus providing
\begin{equation}\label{leading order scalar amplitude}
    \delta\phi^{(0)}(\xi)=\delta\phi^{(0)}(\xi_s)\frac{D(\xi_s)}{\sqrt{\bar{\phi}(\xi_s)}}\left(\frac{\sqrt{\bar{\phi}(\xi)}}{D(\xi)}\right).
\end{equation}

Let us now describe the geometric optics regime for the gravitational sector. The leading-order equation provides the usual dispersion relation $k_{\mu}k^{\mu}=0$.

The equation describing the evolution of the leading order tensor amplitude has a richer structure
\begin{align}\label{eps-1 metric}
\begin{split}
    &\left[\tensor{\mathsf{K}}{_{\mu\nu}^{\rho\sigma\alpha\beta}}
(2k_{\alpha}\nabla_{\beta}+\nabla_{\alpha}k_{\beta})-2k_{\alpha}\tensor{\mathsf{\Tilde{A}}}{_\mu_\nu^\alpha^\beta^\gamma}\delta^{\rho}_{\beta}\delta^{\sigma}_{\gamma}\right]\tilde{h}^{(0)}_{\rho\sigma}=\\&=2\tensor{\mathsf{\Tilde{A}}}{_\mu_\nu^\alpha}k_{\alpha}\delta\phi^{(0)}.
\end{split}
 \end{align}
By plugging the tensor decomposition\til\eqref{amplitude decomposition} into the above relation along with Eqs.\til\eqref{BD effective metric tensor},\til\eqref{eq: A BD} and\til\eqref{eq: Q BD}, and subesequently contracting the result with $\hat{e}^{\mu}_{A}\hat{e}^{\nu}_{B}$, one easily obtains
\begin{align}\label{tensor amplitude evolution BD}
\mathcal{D}\left(\sqrt{\bar{\phi}}\tilde{\alpha}^{(0)}_{AB}\right)= \frac{d}{d\xi}\left[\sqrt{\bar{\phi}}D\tilde{\alpha}^{(0)}_{AB} \right]=0.
\end{align}
It is worth stressing that each amplitude evolves independently and the evolution can be obtained by integrating from the source of the waves to a generic point, $\xi$, thus yielding
\begin{equation}\label{coefficient tensor LO solution}
    \tilde{\alpha}^{(0)}_{AB}(\xi)=\tilde{\alpha}^{(0)}_{AB}(\xi_s)\frac{\sqrt{\bar{\phi}(\xi_s)}D(\xi_s)}{\sqrt{\bar{\phi}(\xi)}D(\xi)}.
\end{equation}
Being the TT gauge conditions valid in such a theory, the constraints on the coefficients $\tilde{\alpha}^{(0)}_{AB}$ and $\tilde{\alpha}^{(1)}_{AB}$ are identically the same as those discussed for GR in Sec.\til\ref{Tetrad decomposition}: in the GO limit only the $mm$ and $ll$ components are non-zero, cf. Eqs.\til\eqref{eq: Leading order tensor amplitude} and\til\eqref{uncontrained tensor modes} hold.
Specifically, by inserting the solution\til\eqref{coefficient tensor LO solution} in Eq.\til\eqref{eq: Leading order tensor amplitude}, one gets the evolution of the leading order tensor amplitude for the Brans-Dicke theory
\begin{equation}\label{LO tensor amplitude BD}
  \tilde{h}^{(0)}_{\mu\nu}(\xi)=\frac{\sqrt{\bar{\phi}(\xi_s)}D(\xi_s)}{\sqrt{\bar{\phi}(\xi)}D(\xi)}\left[\tilde{\alpha}^{(0)}_{mm}(\xi_s)m_{\mu}m_{\nu}+\tilde{\alpha}^{(0)}_{ll}(\xi_s)l_{\mu}l_{\nu}\right].
\end{equation}
The result is similar to that of GR, where the leading-order modes evolve inversely with the distance. However, one can observe a novel modulation factor, $\sqrt{\bar{\phi}(\xi)}$, due to the presence of the scalar field. This is related to the background scalar value modulating the effective Planck's mass, and hence the amplitude of GWs. These results are in complete agreement with Refs.\til\cite{Dalang:2020eaj,Garoffolo:2019mna,Dalang:2019rke}.

\subsubsection{Beyond geometric optics}

The bGO equations are significantly more intricate than that of GR due to additional terms and interactions between the scalar and tensor sectors. This differences lead to novel effects and require a detailed treatment.

For the scalar wave, the NNLO equation captures the bGO correction to the scalar wave,  namely
\begin{align}
\begin{split}
    &\left[\mathsf{\tilde{K}}^{\alpha\beta}\left(2k_{\alpha}\nabla_{\beta}+\nabla_{\beta}k_{\alpha}\right)-k_{\gamma}\mathsf{\Tilde{A}}^{\gamma}\right]\delta\phi^{(1)}=\\&=i\left[\left(\mathsf{\tilde{K}}^{\alpha\beta}\nabla_{\alpha}\nabla_{\beta}+\mathsf{\Tilde{A}}^{\gamma}\nabla_{\gamma}+\mathsf{\Tilde{M}}\right)\delta\phi^{(0)}+\mathsf{\Tilde{M}}_{\phi}^{\alpha\beta}\tilde{h}^{(0)}_{\alpha\beta}\right].
\end{split}
\end{align}
The LHS of the above equation has the same structure of the NLO propagation, Eq.\til\eqref{LO scalar amplitude}, and can be re-written as Eq.\til\eqref{LO conservation Eq}. The main difference is that the RHS introduces a source term, given by the full propagation equation acting on the leading-order scalar amplitude. In other words, the bGO correction is sourced by the failure of the GO prediction to satisfy the full equation, as expected from the general framework discussed in Sec\til\ref{Short wave expasion general}. The general solution for $\delta\phi^{(1)}$ can therefore be expressed as
\begin{widetext}
    \begin{equation}\label{NLO scalar amplitude}
\delta\phi^{(1)}(\xi)=\delta\phi^{(1)}(\xi_s)\frac{D(\xi_s)}{\sqrt{\bar{\phi}(\xi_s)}}\left(\frac{\sqrt{\bar{\phi}(\xi)}}{D(\xi)}\right)+i\frac{\sqrt{\bar{\phi}(\xi)}}{D(\xi)}\int_{\xi_s}^{\xi}d\xi \frac{D(\xi)}{\bar{\phi}^{3/2}}\left[\left(\mathsf{\tilde{K}}^{\alpha\beta}\nabla_{\alpha}\nabla_{\beta}+\mathsf{\Tilde{A}}^{\gamma}\nabla_{\gamma}+\mathsf{\Tilde{M}}\right)\delta\phi^{(0)}+\mathsf{\Tilde{M}}^{\alpha\beta}_{\phi}\tilde{h}^{(0)}_{\alpha\beta}\right].
\end{equation}
\end{widetext}
The result clearly shows how tensor polarizations interact with the scalar sector due to mass-like terms with zero derivatives, $\mathsf{\Tilde{M}}^{\alpha\beta}\tilde{h}^{(0)}_{\alpha\beta}$.
These interactions will source the scalar polarization, even if only tensor modes were emitted.

The tensor equation, at order $\epsilon^{0}$ reads
\begin{equation}
\label{eps0 metric} 
\left[\tensor{\mathsf{K}}{_{\mu\nu}^{\rho\sigma\alpha\beta}}
(2k_{\alpha}\nabla_{\beta}+\nabla_{\alpha}k_{\alpha})-2k_{\alpha}\tensor{\mathsf{\Tilde{A}}}{_\mu_\nu^\alpha^\beta^\gamma}\delta^{\rho}_{\beta}\delta^{\sigma}_{\gamma}\right]\tilde{h}^{(1)}_{\rho\sigma}=i \mathsf{F}_{\mu\nu},
\end{equation}
where
\begin{equation}\label{eq:source_tensor_bGO}
    \begin{split}
\mathsf{F}_{\mu\nu}&\equiv\mathsf{\Tilde{M}}_{\mu\nu}^{h}\delta\phi^{(0)}-k_{\gamma}\tensor{\mathsf{\Tilde{M}}}{_\mu_\nu^\gamma} \delta\phi^{(1)}+\\&+\left[-\frac{1}{2}\tensor{\mathsf{\Tilde{K}}}{_\mu_\nu^\rho^\sigma^\alpha^\beta}\nabla_{\alpha}\nabla_{\beta}+\delta^{\rho}_{\alpha}\delta^{\sigma}_{\beta}\tensor{\mathsf{\Tilde{A}}}{_\mu_\nu^\alpha^\beta^\gamma}\nabla_{\gamma}+\tensor{\mathsf{\Tilde{M}}}{_{\mu\nu}^{\rho\sigma}}\right]\tilde{h}^{(0)}_{\rho\sigma}.        
    \end{split}
\end{equation}
The structure of the LHS of Eq.\til\eqref{eps0 metric} is identical to that of the GO, Eq.\til\eqref{eps-1 metric}, and can be re-written as Eq.\til\eqref{tensor amplitude evolution BD}.
Further, the evolution of the coefficients $\alpha^{(1)}_{AB}$, is obtained by plugging Eq.\til\eqref{amplitude decomposition} into Eq.\til\eqref{eps0 metric} and projecting the result along the dual tetrads basis $\hat{e}^{\mu}_{A}\hat{e}^{\nu}_{B}$, leading to
\begin{equation}
   \frac{d}{d\xi}\left(\tilde{\alpha}^{(1)}_{AB}D\sqrt{\bar{\phi}}\right)=iD(\xi)\hat{e}^{\mu}_{A}\hat{e}^{\nu}_{B}\mathsf{F}_{\mu\nu}.
\end{equation}
By integrating the above equation, one obtains
\begin{equation}
    \begin{split}\label{bgo tensor amplitude BD}
    \tilde{\alpha}_{AB}^{(1)}(\xi)&=\tilde{\alpha}^{(1)}_{AB}(\xi_s)\frac{\sqrt{\bar{\phi}(\xi_s)}D(\xi_s)}{\sqrt{\bar{\phi}(\xi)}D(\xi)}+\\&+\frac{i}{\sqrt{\bar{\phi}(\xi)}D(\xi)}\int_{\xi_s}^{\xi}d\xi' \frac{D(\xi')}{\sqrt{\bar{\phi}}}\left(\hat{e}^{\mu}_{A}\hat{e}^{\nu}_{B}\mathsf{F}_{\mu\nu}\right),
\end{split}
\end{equation}
Note that the $\mathsf{F}$ tensor includes both leading order metric amplitudes and scalar field perturbations, cf. Eq.~\eqref{eq:source_tensor_bGO}.

At this point, we can set without loss of generality $\tilde{\alpha}^{(1)}_{AB}(\xi_s)=\tilde{\alpha}^{(1)}_{AB}(\xi_o)=\delta\phi^{(1)}(\xi_s)=\delta\phi^{(1)}(\xi_o)=0$. It is also reasonable to assume that bGO corrections triggering novel (either apparent or additional) polarizations forbidden in the GO limit do not contribute either at the source or at the observation point, as the lens-induced effects are entirely negligible asymptotically.

In any theory of gravity, mass-like terms, which involve zero-order derivatives, inevitably contribute to wave propagation, making them universal features regardless of the specific framework. In the bGO regime, in particular, the propagating perturbation fields gain a dispersive nature, thus leading to frequency-dependent modifications in the GW signal. Such dispersive effects introduce modifications to both the phase and amplitude of the signal, thus producing corrections that vary with frequency and providing a potential signature that may be detectable across cosmological distances. Importantly, the interactions between the scalar and tensor sectors rely on the presence of an inhomogeneous background; if the propagation medium were homogeneous, such as in a cosmological setting, the interactions would effectively vanish. Consequently, the presence of a gravitational lens or a similar inhomogeneous structure is necessary for these effects to manifest.

\section{Point-like lens}\label{Point like lens}

\begin{figure*}[htbp]
    \centering
    \begin{minipage}[t]{0.45\textwidth}
        \centering
        \begin{tikzpicture}
        \draw[->] (-4,0)--(4,0) node[right]{$\hat{z}$};
        \draw[->] (0,-.5)--(0,3.5) node[above]{$\hat{x}$};
        \draw[<->,opacity=1, ceruleanblue] (0,0) -- (0,2.7);
        \node at (-.3,2.4) [below] {\color{ceruleanblue} $b$};

        \draw[loosely dotted] (-3.8,0)--(-3.8,2.7) (-3.8,0) node[below]{$z_s$};
        \draw[loosely dotted] (3.8,0)--(3.8,2.7) (3.8,0) node[below]{$z_o$};
        \draw[dashdotted] (-4,2.7)--(4,2.7);

        \draw[domain=-4:4, Magenta,opacity=0.75] plot (\x, {-0.012*(\x)^2+2.7});

        \fill (-3.9,2.8) circle (.18);
        \fill (-3.7,2.5) circle (.1);

        \tikzset {myCircle/.style= {ceruleanblue, path fading=fade out, opacity=0.88}}
        \shade[myCircle,inner color=Cyan, outer color=Thistle, fill opacity=.88] (0,0) circle (1.6);
        \node [black,scale=1.2] at (0,0,0) {$\bigstar$};

        \coordinate (A) at (3.7,2.35);
        \coordinate (B) at (4.2,2.5);
        \coordinate (C) at (3.8,2.9);
        \draw[red] (A) -- (B) -- (C) -- cycle;
        \fill (A) circle (2pt);
        \fill (B) circle (2pt);
        \fill (C) circle (2pt);

        \node[above] at (-3.8,3.9) {\text{Source plane}};
        \node[above] at (0,3.9) {\text{Lens plane}};
        \node[above] at (3.8,3.9) {\text{Observer plane}};
            \node[below] at (0,-1.8) {\text{Longitudinal view}};
    \end{tikzpicture} 

    \end{minipage}
    \hfill
    \begin{minipage}[t]{0.45\textwidth}
        \centering
        \begin{tikzpicture}

    \draw[->] (-4,0) -- (4,0) node[right] {$\hat{y}$};
    \draw[->] (0,-.5) -- (0,3.5) node[above] {$\hat{x}$};

    \tikzset {myCircle/.style= {ceruleanblue, path fading=fade out, opacity=0.88}}
    \shade[myCircle,inner color=Cyan, outer color=Thistle, fill opacity=.88] (0,0) circle (1.6);
    \node [black,scale=1.2] at (0,0,0) {$\bigstar$};

    \draw[-] (0,0) -- (2.3,2.3) node[midway, above] {$r$};
\draw[] (2.3,2.3) circle (0.2); 
\draw[rotate around={45:(2.3,2.3)}] (2.3,2.5) -- (2.3,2.1);
\draw[rotate around={45:(2.3,2.3)}] (2.1,2.3) -- (2.5,2.3);  

\node[above] at (2.3,2.7) {Entering wave ($\hat{z}$)};

    \draw[] (0,0.7) arc[start angle=90, end angle=45, radius=0.7];
    \draw[<->,opacity=1, ceruleanblue] (0,0) -- (0,2.3);
    \node at (0.5,1.0) {$\beta$};

    \draw[loosely dotted] (2.3,2.3) -- (2.3,0) node[below] {$r\sin\beta$};
    \draw[loosely dotted] (2.3,2.3) -- (0,2.3) node[left] {$b = r\cos\beta$};
    \node[below] at (0,-1.8) {\text{Transverse view (lens plane)}};

\end{tikzpicture}
    \end{minipage}
    \caption{Schematic representation of the propagation process for scalar and tensor waves. The \textbf{left panel} shows the $(x,z)$ plane, illustrating the wave propagation along the $\hat{z}$-axis. The point-lens located at the origin (black star) and its surrounding scalar field (depicted as a halo) cause a weak deflection of the wave trajectory, represented by the solid purple line. In our case of interest, $z_s \rightarrow -\infty$ and $z_o \rightarrow +\infty$. The \textbf{right panel} illustrates the lens plane ($x$, $y$), where the impact position of scalar and tensor waves signal is identified by the angle $\beta$. In our calculations, we set $\beta = 2\pi$, effectively eliminating the component along the $\hat{y}$-axis.}
    \label{fig:propagation scheme}
\end{figure*}
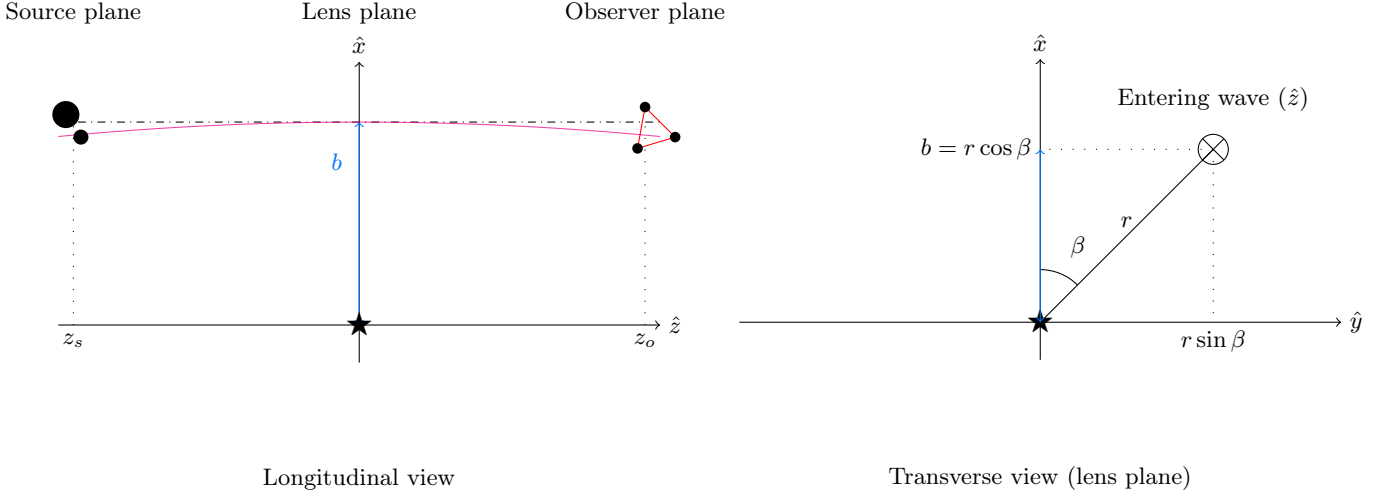

In this section, we apply the formalism to the  point-like lens case, discussing the bGO dispersive corrections to the scalar and tensor amplitudes in Brans-Dicke theory.
In presence of a point-like lens, the line element reads\footnote{In this work, corrections to the Jordan gravitational potential have not been considered.}
\begin{equation}\label{wfl metric}
    ds^2=-(1+2\Psi)dt^2+(1-2\Psi)(dx^2+dy^2+dz^2),
\end{equation}
where $\Psi\equiv -R_s/2R$ is the gravitational potential such that $\Psi\ll1$,  $R_s\equiv 2GM_L$ is the Schwarzchild radius with $M_L$ as the lens' mass and $R\equiv\sqrt{x^2+y^2+z^2}$ the radial distance from the origin of the coordinate system, in which the lens is located (See Fig.\til\ref{fig:propagation scheme}).
We will follow the perturbative approach, up to linear order in $\Psi$, presented in Ref.\til\cite{Dalang:2021qhu}.
The first step is to expand the tensor amplitude\til\eqref{amplitude decomposition} to the first order in the gravitational potential, yielding
\begin{equation}\label{perturbed amplitudes WFL}
\tilde{h}^{(n)}_{\mu\nu}=\bar{\tilde{\alpha}}^{(n)}_{AB}\bar{e}^A_{(\mu}\bar{e}^B_{\nu)}+2\bar{\tilde{\alpha}}^{(n)}_{AB}\delta e^{A}_{(\mu}e^{B}_{\nu)}+\delta \tilde{\alpha}^{(n)} _{AB}\bar{e}^A_{(\mu}\bar{e}^B_{\nu)},
\end{equation}
where the bar identifies the background quantities constructed on the Minkowski metric, and the variations encode the effects of the gravitational potential characterizing the lens. 

We will consider propagation along the $\hat{z}$-axis in the $(x,z)$-plane, by setting the impact angle ($\beta$) on the lens plane as $\beta=2\pi$, $i.e.$ $y=0$ (See Fig.~\ref{fig:propagation scheme}, right panel). 
Specifically, given that $\Psi \ll 1$, the expected deflection angle between the lens-perturbed trajectory. %
Therefore, we can compute the deflection as leading-order corrections to the unperturbed trajectory (Born approximation) and choose the affine parameter to be $d\xi = dz/\Omega$, where $\Omega\equiv2\pi/\lambda$ is a constant amplitude for the $4-$vector momentum and $\lambda$ is the signal's wavelenght (see Refs.\til\cite{Cusin:2019rmt, Dalang:2021qhu}).
The background tetrads can be constructed to be constants
\begin{align}\label{Background tetrads1}
   & \bar{k}^{\mu}\equiv\Omega(1,0,0,1),\\& 
   \label{Background tetrads2}
   \bar{n}^{\mu}\equiv\frac{1}{2\Omega}(1,0,0,-1),\\ 
   \label{Background tetrads3}
  & \bar{m}^{\mu}\equiv\frac{1}{\sqrt{2}}(0,1,i,0),\\&\label{Background tetrads4}\bar{l}^{\mu}\equiv\frac{1}{\sqrt{2}}(0,1,-i,0).
\end{align}

Further, it is also straightforward to verify that this construction fulfills all the properties highlighted in Sec.\til\ref{Tetrad decomposition}.
The perturbation to the tedrads at the first order in $\Psi$ can be explicitly evaluated by integrating the linearized geodesic equation given by\til\cite{Cusin:2017fwz}
\begin{align}
\begin{split}
    \delta e^{\mu}_A&=-\int_{z_s}^{z_o
    }\frac{dz}{\Omega}\delta\Gamma^{\mu}_{\alpha\beta}\bar{e}^{\alpha}_A\bar{k}^{\beta}=-\int_{-\infty}^{+\infty
    }\frac{dz}{\Omega}\delta\Gamma^{\mu}_{\alpha\beta}\bar{e}^{\alpha}_A\bar{k}^{\beta},
\end{split}
\end{align}
where, in the second equality, we restricted to the special case in which the lens is far away from both the source and the observer $z_s\rightarrow -\infty$, $z_o \rightarrow +\infty$. Here $\delta\tensor{\Gamma}{^{\mu}_\alpha_\beta}$ are the linerized Christoffel symbols, evaluated using Eq.\til\eqref{wfl metric} (whose expressions are shown in Appendix\til\ref{appendix: Weak field limit}), up to the first order in $\Psi$. The subscripts $s$ and $o$ denote the points in which the source and the observer are located, respectively. By plugging Eqs.\til(\ref{Background tetrads1}--\ref{Background tetrads4}) into the above integral, one gets
\begin{align}
    k^{\mu}&=\bar{k}^{\mu}+\delta k^{\mu}=\Omega\left(1,-\frac{2R_s}{b},0,1\right),\\
     n^{\mu}&=\bar{n}^{\mu}+\delta n^{\mu}=\frac{1}{2\Omega}\left(1,0,0,-1\right),\\
      m^{\mu}&=\bar{m}^{\mu}+\delta m^{\mu}=\frac{1}{\sqrt{2}}\left(-\frac{R_s}{b},1,i,\frac{R_s}{b}\right),\\
       l^{\mu}&=\bar{l}^{\mu}+\delta l^{\mu}=\frac{1}{\sqrt{2}}\left(-\frac{R_s}{b},1,-i,\frac{R_s}{b}\right),
\end{align}
where $b\equiv\sqrt{x^2+z^2}$ is the impact parameter.

\subsection{General relativity}\label{GR point lens}
Let us first analyze how the GO amplitude\til\eqref{Covariant Tensor amplitude GR final} is modified at first order in $\Psi$. In particular, one has
\begin{equation}\label{leading order perturbed tensor amplitude}
\begin{split}
    \tilde{h}^{(0)}_{\mu\nu}(z_o)=&\frac{\bar{D}(z_s)}{\bar{D}(z_o)}\left[\tilde{\alpha}^{(0)}_{mm}(z_s)\bar{m}_{\mu}\bar{m}_{\nu}+\tilde{\alpha}^{(0)}_{ll}(z_s)\bar{l}_{\mu}\bar{l}_{\nu}\right]\Delta(z_o)+\\
    +&\frac{\bar{D}(z_s)}{\bar{D}(z_o)}\left[2\tilde{\alpha}^{(0)}_{mm}(z_s)\bar{m}_{(\mu}\delta m_{\nu)}+2\tilde{\alpha}^{(0)}_{ll}(z_s)\bar{l}_{(\mu}\delta l_{\nu)}\right],
\end{split}
\end{equation}
with
\begin{equation}
    \Delta(z_o)\equiv\left(1-\frac{\delta D(z_o)}{\bar{D}(z_o)}\right),
\end{equation}
where $\delta D$ represents the first-order perturbation in the comoving distance along the geodesic\til\cite{Bonvin:2005ps}; the latter can be neglected at the first order in $\Psi$ (as shown in Ref.\til\cite{Dalang:2021qhu}, Appendix A). Moreover, for simplicity, we assumed a Euclidean distance parameterized as $\bar{D}(z) = z - z_s$.
Recalling Eq.\til\eqref{perturbed amplitudes WFL}, the correction to the tensor amplitude $\tilde{h}^{(1)}_{\mu\nu}$, at the linear order in $\Psi$, is 
\begin{align}\label{perturbed bgo tensor amplitude}
    \begin{split}
     \tilde{h}^{(1)}_{\mu\nu}(z_o)&=\delta\tilde{\alpha}^{(1)}_{nm}(z_o)\bar{n}_{(\mu}\bar{m}_{\nu)}+\delta\tilde{\alpha}^{(1)}_{nl}(z_o)\bar{n}_{(\mu}\bar{l}_{\nu)}+\\&+\delta\tilde{\alpha}^{(1)}_{mm}(z_o)\bar{m}_{\mu}\bar{m}_{\nu}+\delta\tilde{\alpha}^{(1)}_{ll}(z_o)\bar{l}_{\mu}\bar{l}_{\nu}+\\&+\delta\tilde{\alpha}^{(1)}_{nn}(z_o)\bar{n}_{\mu}\bar{n}_{\nu}.   
    \end{split}
\end{align}

In GR, the only coefficient surviving is $\delta\tilde{\alpha}^{(1)}_{nn}$, discussed in Ref.\til\cite{Dalang:2021qhu}, thus providing
\begin{equation}\label{GR nn mode}
   \tilde{h}^{(1)}_{\mu\nu}(z_o) = -i\frac{\bar{D}(z_s)}{\bar{D}(z_o)}\frac{4\Omega R_s}{b^2}\tilde{\zeta}^{(0)}_{+}(z_s)\Bar{n}_{\mu}\Bar{n}_{\nu},
\end{equation}
where we defined the quantities
\begin{align}
    \tilde{\zeta}^{(0)}_{+}(z_s)&\equiv\tilde{\alpha}^{(0)}_{mm}(z_s)+\tilde{\alpha}^{(0)}_{ll}(z_s),\\
    \tilde{\zeta}^{(0)}_{-}(z_s)&\equiv\tilde{\alpha}^{(0)}_{mm}(z_s)-\tilde{\alpha}^{(0)}_{ll}(z_s).
\end{align}
The latter are nothing but symmetric and antisymmetric combinations of the left- and right-handed modes contributing to the GO tensor amplitude. The above definitions can be further related to standard GW polarization via Eqs.\til\eqref{eq:hpluscross_alpha1} and\til\eqref{eq:hpluscross_alpha2}, thus yielding
\begin{align}
    \tilde{\zeta}^{(0)}_{+}(z_s)&=\sqrt{2}\tilde{h}^{(0)}_{+}(z_s),\\
    \tilde{\zeta}^{(0)}_{-}(z_s)&=-i\sqrt{2}\tilde{h}^{(0)}_{\times}(z_s).
\end{align}

Note that Eq.\til\eqref{GR nn mode} does not have components along $m_{\mu}m_{\nu}$ and $l_{\mu}l_{\nu}$, implying that there are no dispersive corrections affecting the left- and right-handed polarizations. However, the coupling between GWs and the background curvature induced by the lens triggers the existence of a new apparent polarization. 
In particular, this effect actually arises from the failure of the left- and right-handed polarizations to remain orthogonal to the wave's propagation direction.

Recalling also Eq.\til\eqref{Background tetrads2}, Eq.\til\eqref{GR nn mode} shows an overall factor of $1/\Omega$, thus suggesting an explicit frequency dependence, which scales as $1/f$ as expected for bGO corrections leading to LID.
In the high-frequency limit, $\Omega \rightarrow \infty$, these corrections vanish identically. %

\subsection{Massless Brans-Dicke}\label{PLL Brans Dicke}

In this section, we discuss the bGO corrections for the massless Brans-Dicke theory. 
For transparency, we will restrict our analysis to a simple spherically symmetric background scalar field configuration:\til\cite{Faraoni:2009km}
\begin{equation}\label{Spherically symmetric background scalar field}
\bar{\phi}(R)\equiv\bar{\phi}_{\infty}+q\frac{ G M_L}{R},
\end{equation}
where $\bar{\phi}_{\infty}$ is an asymptotic costant value as $R\rightarrow\infty$ and $q$ a dimensionless parameter. Furthermore, given its constant nature, we can set $\phi_{\infty}=1$, thus recovering GR at the asymptotic limit when $R$ is large.

Let us now start by analyzing the contribution of bGO arising from the scalar sector. By evaluating Eq.\til\eqref{NLO scalar amplitude} using the BD matrices coefficients shown in Appendix\til\ref{Terms Brans Dicke} along with Eqs.\til\eqref{leading order scalar amplitude},\til\eqref{leading order perturbed tensor amplitude}, and\til\eqref{Spherically symmetric background scalar field}, the scalar amplitude correction can be re-written in a more compact form as
\begin{align}\label{scalar amplitude correction}
    \begin{split}
    \delta\phi^{(1)}(z_o)=&i\frac{\sqrt{\bar{\phi}(z_o)}}{\bar{D}(z_o)} \int_{-\infty}^{+\infty}\frac{dz}{\Omega}\left[f^{\phi}_{\infty}(b,\tau;z)\delta\phi^{(0)}(z_s)\right.+\\&\left.+f^{+}_{\infty}(b,\tau;z)\tilde{\zeta}_{+}^{(0)}(z
    _s)+f^{-}_{\infty}(b,\tau;z)\tilde{\zeta}_{-}^{(0)}(z_s)\right],
    \end{split}
\end{align}
where $\tau\equiv q G M_L$ and the subscript $\infty$ refers to the form of the functions when the limit $z_s\rightarrow-\infty$ is performed.%
\footnote{In general, $f^{\phi,\pm}$ show an explicit dependence on $z_s$ as well (refer to Appendix\til\ref{Function of the modes})
Note because the limit $z_s\rightarrow-\infty$ and $z_o\rightarrow+\infty$, 
affects not only the integration range, but also the integrand $f^{\phi,\pm}(b,z_s,\tau;z) \to f^{\phi,\pm}_{\infty}(b,\tau;z)$. It can be verified that the limit is well behaved.}

Moreover, the superscript denotes the GO variable to which the function is associated ($e.g.$, $f^{\phi}$ is the function associated with $\delta\phi^{(0)}$, $f^{\pm}$ to $\tilde{\zeta}_{\pm}^{(0)}(\xi_s)$, respectively). The functions $f^{\phi,\pm}_{\infty}(b,\tau;z)$ are presented in the Appendix\til\ref{Function of the modes}. 

The result shows that the dispersive scalar correction is sourced by GO amplitudes, $\delta\phi^{(0)}(z_s)$ and $\tilde{\zeta}_{\pm}^{(0)}(z_s)$, thus encoding $\tilde{\alpha}^{(0)}_{mm}(z_s)$ and $\tilde{\alpha}^{(0)}_{ll}(z_s)$.
The integration of the aforementioned equation can be performed analytically, thus obtaining
\begin{align}\label{eq:bGO scalar PLL}
\begin{split}
        \delta\phi^{(1)}(z_o)&=\frac{i}{\Omega}\frac{\sqrt{\bar{\phi}(z_o)}}{\bar{D}(z_o)}\left[F^{\phi}(b,\tau)\delta\phi^{(0)}(\xi_s)+\right.\\&\left.+F^{+}(b,\tau)\tilde{\zeta}^{(0)}_{+}(z_s)+F^{-}(b,\tau)\tilde{\zeta}^{(0)}_{-}(z_s)\right],
\end{split}
\end{align}

where $F^{\phi,\pm}(b,\tau)$ are the resulting integrals of the corresponding functions $f^{\phi,\pm}_{\infty}(b,\tau;z)$ whose explicit form is shown in the Appendix\til\ref{Analytic integrals}.
It can be easily observed that such a correction is a frequency-dependent quantity, due to the presence of $\Omega$, which behaves as $1/f$ and contributes as a modification to the phase of the scalar wave, due to the presence of the imaginary unit. The high-frequency limit causes this correction to vanish.

Let us derive the bGO corrections to the leading-order tensor amplitude by computing the coefficients appearing in Eq.\til\eqref{perturbed bgo tensor amplitude}. By perturbing Eq.\til\eqref{bgo tensor amplitude BD} at the first order in $\Psi$ along with the BD matrix coefficients appearing in Appendix\til\ref{Terms Brans Dicke} to make $\mathsf{F}_{\mu\nu}$ explicit in terms of the metric and the scalar field, one gets
\begin{widetext}\begin{align}\label{eq: perturbed modes general eq}
\begin{split}
    \delta\tilde{\alpha}^{(1)}_{AB}(z_o)= \frac{i}{\sqrt{\Bar{\phi}(z_o)}\bar{D}(z_o)}\int_{-\infty}^{+\infty}\frac{dz}{\Omega}\hat{e}^{\alpha}_{A}\hat{e}^{\beta}_{B} \frac{\bar{D}(z)}{\sqrt{\bar{\phi}}}&\left\{-2(\delta R_{\alpha\beta})\delta\phi^{(0)}-2\delta \tilde{h}^{(0)}_{\alpha\beta}\Box\bar{\phi}+\frac{2\omega X\delta \tilde{h}^{(0)}_{\alpha\beta}}{\bar{\phi}}-\frac{2(\delta\tensor{\Gamma}{^\gamma_{\alpha\beta}})\bar{\phi}_{\gamma}\delta\phi^{(0)}}{\bar{\phi}}+\right.\\
    &\left.+2\delta\tensor{\Gamma}{^\rho_{\alpha\beta}}\tilde{h}^{(0)}_{\gamma\rho}\bar{\phi}^{\gamma}+2\delta\tensor{\Gamma}{^\rho_{\gamma[\beta}}\tilde{h}^{(0)}_{\alpha]\rho}\bar{\phi}^{\gamma}-2\tilde{h}^{(0)}_{\alpha\beta}\delta\tensor{\Gamma}{^\rho_{\gamma\rho}}\bar{\phi}^{\gamma}-2\partial_{\beta}(\delta\tilde{h}^{(0)}_{\alpha\gamma})\bar{\phi}^{\gamma}+\right.\\&\left.+\partial_{\gamma}(\delta\tilde{h}^{(0)}_{\alpha\beta})\bar{\phi}^{\gamma}
+\bar{\phi}\left[(\delta R) \tilde{h}^{(0)}_{\alpha\beta}-2(\delta\tensor{R}{^\gamma_\beta})\tilde{h}^{(0)}_{\alpha\gamma}-2(\delta\tensor{R}{_{\gamma\alpha\beta\rho}})\tilde{h}^{(0){\gamma\rho}}+\right.\right.\\&+\Box(\delta\tilde{h}^{(0)}_{\alpha\beta})-4\delta\tensor{\Gamma}{^\gamma_{\rho(\alpha}}\partial^{\rho}\tilde{h}^{(0)}_{\beta)\gamma}-2\partial^{\rho}(\delta\tensor{\Gamma}{^\gamma_{\rho(\alpha}})\tilde{h}^{(0)}_{\beta)\gamma}+\delta\tensor{\Gamma}{^\gamma_{\gamma\rho}}\partial^{\rho}\tilde{h}^{(0)}_{\alpha\beta}\left.\left.\right]\right\}.
\end{split}
\end{align}
\end{widetext}

Subsequently plugging Eqs.\til\eqref{leading order scalar amplitude},\til\eqref{leading order perturbed tensor amplitude}, and\til\eqref{Spherically symmetric background scalar field} into the above relation, 
each perturbed mode in the null tetrad basis can be expressed as
\begin{align}\label{perturbed tensor in z}
\begin{split}
            \delta\tilde{\alpha}^{(1)}_{AB}(z_o)=&\frac{i}{\sqrt{\Bar{\phi}(z_o)}\Bar{D}(z_o)}\int_{-\infty}^{+\infty}\frac{dz}{\Omega}\times\\&\times\left[f^{\phi}_{AB_{\infty}}(b,\tau;z)\delta\phi^{(0)}(z_s)+\right.\\&\left.+f^{+}_{AB_{\infty}}(b,\tau;z)\tilde{\zeta}^{(0)}_{+}(z_s)\right.+\\&\left.+f^{-}_{AB_{\infty}}(b,\tau;z)\tilde{\zeta}^{(0)}_{-}(z_s)\right],
\end{split}
\end{align}
where we refer to Appendix\til\ref{Function of the modes} for  $f^{\phi,\pm}_{AB_{\infty}}(b,\tau;z)$. The subscript $AB$ indicates that these functions contain, in their definition, the contraction with the appropriate dual tetrad $\hat{e}^{\alpha}_{A}\hat{e}^{\beta}_{B} $. Similarly to the scalar case, the result can be written in terms of leading-order scalar amplitude and tensor modes

\begin{widetext}
    \begin{align}\label{eq: tensor bgo corrections}
\begin{split}
\begin{pmatrix}
\delta\tilde{\alpha}^{(1)}_{mm}(z_o) \\
\delta\tilde{\alpha}^{(1)}_{ll}(z_o)\\
\delta\tilde{\alpha}^{(1)}_{nn} (z_o)\\
\end{pmatrix}
=\frac{i}{\sqrt{\bar{\phi}(z_o)}\Bar{D}(z_o)}
\begin{bmatrix}
F^{\phi}_{mm} (b,\tau)& F^{+}_{mm} (b,\tau)& F^{-}_{mm}(b,\tau)\\
F^{\phi}_{ll}(b,\tau) & F^{+}_{ll} (b,\tau)& F^{-}_{ll}(b,\tau) \\
F^{\phi}_{nn}(b,\tau) & F^{+}_{nn}(b,\tau) & 0\\
\end{bmatrix}
\begin{pmatrix}
\delta\phi^{(0)}(z_s) \\
\tilde{\zeta}^{(0)}_{+}(z_s)\\
\tilde{\zeta}^{(0)}_{-}(z_s).
\end{pmatrix},
\end{split}
\end{align}
\begin{align}
    \delta\tilde{\alpha}^{(1)}_{nm}(z_o)=\delta\tilde{\alpha}^{(1)}_{nl}(z_o)=0.
\end{align}
\end{widetext}
where the functions $F^{\phi,\pm}_{AB}(b,\tau)$ are shown in Appendix\til\ref{Analytic integrals}. The above equation, together with\til\eqref{eq:bGO scalar PLL}, describes the bGO corrections in the massless BD theory.

The existence of scalar radiation and the non-minimal coupling of the scalar field with curvature trigger, from Eq.\til\eqref{eq: perturbed modes general eq}, $\delta\tilde{\alpha}^{(1)}_{mm}$ and $\delta\tilde{\alpha}^{(1)}_{ll}$ to become new non-zero coefficients. Note that these terms arise exclusively in the BD scenario and are completely absent in the GR case (refer to table\til\ref{fig:table polarizations} for a summary). Recalling that the $mm$ and $ll$ modes are related to the $ + $ and $ \times$ polarizations, it is important to emphasize that these corrections have a direct impact on the GW polarizations. 

Secondly, the $\delta\tilde{\alpha}^{(1)}_{nn}$-mode emerge as an apparent scalar polarization, and while it is also predicted by GR, it exhibits distinct differences in comparison. In the GR case, one can verify that most of the integrals in Eq.\til\eqref{eq: perturbed modes general eq} vanish and the only term yielding a non-zero outcome is $(\delta\tensor{R}{_{\gamma\alpha\beta\rho}})\tilde{h}^{(0){\gamma\rho}}$, thus providing the result shown in Eq.\til\eqref{GR nn mode}.\footnote{To study the GR case from Eq.\til\eqref{eq: perturbed modes general eq}, one simply needs to discard the contributions from the scalar field and the scalar wave by setting $\bar{\phi} = \text{constant}$. Consequently, the derivatives of the field will vanish, and the amplitudes at every order of the WKB expansion will be zero.}
In BD instead, it can be observed that there are two additional contributions from Eq.\til\eqref{eq: perturbed modes general eq}, namely $(\delta\tensor{\Gamma}{^\gamma_{\alpha\beta}})\bar{\phi}_{\gamma}\delta\phi^{(0)}/\bar{\phi}$ and $\delta\tensor{\Gamma}{^\rho_{\alpha\beta}}\tilde{h}^{(0)}_{\gamma\rho}\bar{\phi}^{\gamma}$.

The first term is related to the existence of scalar waves that influence the evolution of tensor amplitude, whereas the second one pertains to the coupling between scalar fields and the gravitational dynamics. Essentially, this means that even if we disregard scalar waves and focus exclusively on how the background scalar field impacts the propagation of tensor waves, there would nevertheless appear a discrepancy from what GR results predict. Note that $\delta\tilde{\alpha}^{(1)}_{nm}(z_o)$ and $\delta\tilde{\alpha}^{(1)}_{nl}(z_o)$ are vanishing even for BD. In general, the limit to GR is straightforward to perform by setting $\delta\phi^{(0)},\tau,\beta \to 0$.

In BD, the bGO regime not only induces a new apparent polarization (with novel contributions due to the existence of the scalar wave and scalar-gravitational coupling) but also leads to corrections along the left- and right-handed polarizations, thus producing additional phase modifications to the GWs phase. LID corrections to standard GW polarizations are a smoking gun of BD: Detecting a non-zero $\delta\tilde{\alpha}^{(1)}_{mm}$ and $\delta\tilde{\alpha}^{(1)}_{ll}$ would be a clear signature of deviations from GR.

Finally, once all coefficients are plugged into Eq.\til\eqref{perturbed bgo tensor amplitude}, along with Eqs.\til(\ref{Background tetrads1}--\ref{Background tetrads4}), the result will account for an overall bGO dispersive correction / LID scaling as $1/\Omega$, thus as $1/f$.

Summarizing, we derived analytical expressions for the LID corrections for a point-like lens. We considered a spherically symmetric background scalar field, performing the calculation perturbatively up to the first order in the gravitational potential, assuming a Euclidean comoving distance on the geodesic. We restricted the analysis to the special case in which source and observation points are pushed to infinity, $i.e.$ $z_s\rightarrow-\infty, z_o\rightarrow+\infty$. This situation in BD triggers the $nn$ component, similarly to GR but with a different amplitude and dependence on the parameters. This can be interpreted as an apparent polarization caused by the fact that, in the bGO regime, the polarizations are no longer orthogonal to the propagation direction. Additionally, there are corrections to the \( mm \) and \( ll \) components, which directly impact the standard GW polarizations: these corrections are absent in GR and provide a smoking gun for BD or other alternative theories.

\subsection{Jordan frame}\label{subsec:jordan-frame}
The above results were obtained in the Einstein frame, in which the dynamical variables decouple. To connect with observables, we need to rewrite them in the Jordan frame, in which matter couples minimally to the metric perturbation.
In order to do that, we need to reverse Eq.\til\eqref{BD diagonalization transformation} by obtaining
\begin{equation}
    h_{\mu\nu}=\tilde{h}_{\mu\nu}-\frac{1}{2}g_{\mu\nu}\tilde{h}-g_{\mu\nu}\frac{\delta\phi}{\bar{\phi}}.
\end{equation}
By recalling $\tilde{h}=0$, the above equation reduces to
\begin{equation}
     h_{\mu\nu}=\tilde{h}_{\mu\nu}-g_{\mu\nu}\frac{\delta\phi}{\bar{\phi}}.
\end{equation}
We note that the short-wave approximation\til\eqref{WKB decomposition metric} holds for $h_{\mu\nu}$ as well. By plugging the latter ansatz into the above equation, the relation on the amplitude, at any order $n$, yields
\begin{equation}\label{eq: Jordan Einstein frame amplitude general}
    h^{(n)}_{\mu\nu}=\tilde{h}_{\mu\nu}^{(n)}-g_{\mu\nu}\frac{\delta\phi^{(n)}}{\bar{\phi}}.
\end{equation}
Applying the general rank-2 tensor decomposition\til\eqref{amplitude decomposition} on both sides of Eq.\til\eqref{eq: Jordan Einstein frame amplitude general}, one gets the general transformation of the expansion coefficients between the Einstein and Jordan frames  
\begin{equation}
\alpha_{CD}^{(n)}\Theta_{\mu\nu}^{AB}=\tilde{\alpha}^{(n)}_{AB}\Theta_{\mu\nu}^{AB}-g_{\mu\nu}\frac{\delta\phi^{(n)}}{\bar{\phi}}.
\end{equation}

At the leading order in the short-wave expansion, by using Eq.\til\eqref{eq:metric_tensor_decomposition}, employing the appropriate gauge conditions, in both Jordan and Einstein frames, and subsequently contracting the above result with dual tetrads, one gets

\begin{align}
\alpha^{(0)}_{mm} &= \tilde{\alpha}^{(0)}_{mm}, \\
\alpha^{(0)}_{ll} &= \tilde{\alpha}^{(0)}_{ll}, \\\label{eq:scalar_sourced_mode_0}
\alpha^{(0)}_{ml} &= -\frac{\delta\phi^{(0)}}{\bar{\phi}},
\end{align}
with all the other coefficients vanishing.

Coefficients describing bGO corrections are evaluated in the same way. The nonzero LID terms read%
\begin{align}
\alpha^{(1)}_{mm} &= \tilde{\alpha}^{(1)}_{mm}, \\
\alpha^{(1)}_{ll} &= \tilde{\alpha}^{(1)}_{ll}, \\
\alpha^{(1)}_{nn} &= \tilde{\alpha}^{(1)}_{nn}, \\
\label{eq:scalar_sourced_mode_1}
\alpha^{(1)}_{ml} &= -\frac{\delta\phi^{(1)}}{\bar{\phi}}\,,
\end{align}
while $\alpha^{(1)}_{nm} = \tilde{\alpha}^{(1)}_{nm}$ and $\alpha^{(1)}_{nl} = \tilde{\alpha}^{(1)}_{nl}$ vanish identically.
Moreover, as the bGO corrections are derived at first order in the gravitational potential $\Psi$, the aforementioned relations apply to the terms $\delta\tilde{\alpha}^{(1)}_{AB}$ by replacing $\tilde{\alpha}^{(1)}_{AB}$ with $\delta\tilde{\alpha}^{(1)}_{AB}$.

From the above equations, it can be observed that the scalar field perturbation excites the $ml$-mode, while $mm$, $ll$, and $nn$ coefficients remain invariant under the mapping between the Einstein and Jordan frames. Ultimately, we can express the GO and bGO tensor amplitudes in the Jordan frame as
\begin{align}\label{eq:GO-JF}
    h^{(0)}_{\mu\nu}&=\alpha^{(0)}_{mm}\bar{m}_{\mu}\bar{m}_{\nu}+\alpha^{(0)}_{ll}\bar{l}_{\mu}\bar{l}_{\nu}-\frac{\delta\phi^{(0)}}{\bar{\phi}}\bar{m}_{(\mu}\bar{l}_{\nu)},\\ \label{eq:bGO-JF}
    h^{(1)}_{\mu\nu}&=\delta\alpha^{(1)}_{mm}\bar{m}_{\mu}\bar{m}_{\nu}+\delta\alpha^{(1)}_{ll}\bar{l}_{\mu}\bar{l}_{\nu}+\delta\alpha^{(1)}_{nn}\bar{n}_{\mu}\bar{n}_{\nu}-\frac{\delta\phi^{(1)}}{\bar{\phi}}\bar{m}_{(\mu}\bar{l}_{\nu)}.
\end{align}

\subsection{Phase correction to tensor waves}

\begin{figure*}
  \begin{minipage}[h!]{\textwidth}
{\includegraphics[width=\linewidth]{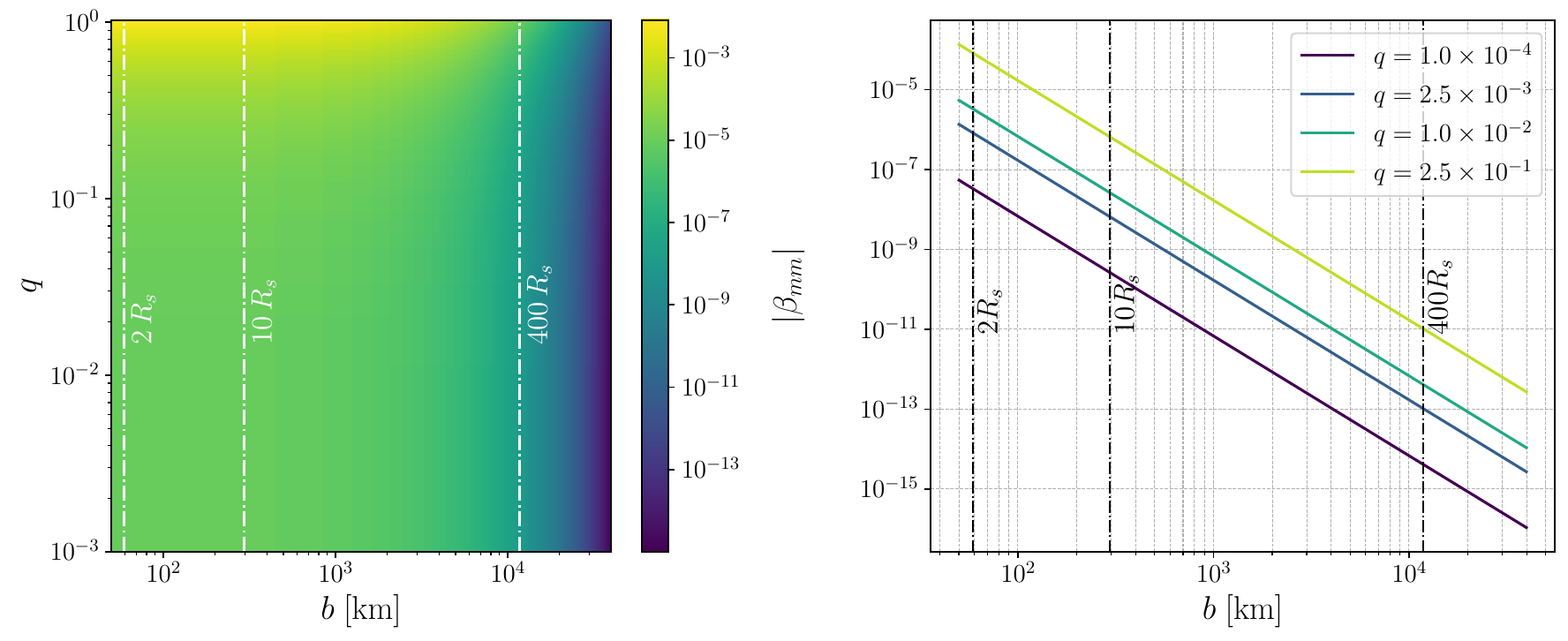}\label{fig:subfig2}}\hfill

  \end{minipage}
      \caption{Dependence of $\beta_{mm}$ on the impact parameter $b$ and scalar charge $q$, for a fixed lens mass $M_L = 10 M_\odot$. \textbf{Left panel:} Heatmap illustrating the variation of $|\beta_{mm}|$ across different $b$ and $q$ values, with key distances marked in units of Schwarzschild radii ($R_s$). \textbf{Right panel:} Line plots showing the sensitivity of $|\beta_{mm}|$ with respect to the impact parameter, for different $q$ values. The effects are most significant for large $q$ values and distances near $2-5 R_s$, while rapidly dropping around $10 R_s$. The plots are shown for the $mm$ mode but are also valid for the $ll$ mode. 
      }
  \label{fig:beta}
\end{figure*}

We now estimate the impact of bGO phase corrections on the waveform, neglecting the scalar polarization for simplicity. We focus solely on the analysis of the $mm$ and $ll$ modes (relevant for $h_+$ and $h_\times$), whose bGO corrections are described by Eq.~\eqref{eq: tensor bgo corrections}, and specifically by Eq.~\eqref{eq:mm_ll_bgo_corrections}.

The bGO phase is characterized by the dimensionless parameter $\beta_{J}$, where $J=\{mm, ll\}$, introduced in Sec.~\ref{sec:general_theory}, Eq.~\eqref{eq:beta_phase} (and discussed in more detail in Refs.~\cite{Oancea:2022szu,Oancea:2023hgu}). The latter provides a measure of the magnitude of the correction to the GW waveform phase, as illustrated in the example of Fig.~\ref{fig:waveforms}. In the case of Brans-Dicke theory, $\beta_{J}=\beta_{J}(b, M_L, \tau)$ is a function of the lens mass, impact parameter, and $\tau\equiv q G M_L$, which encodes the effect of the scalar charge $q$ in its definition. 

We focus on the effects on $\beta_{mm}$, which are equally valid for $\beta_{ll}$ since these quantities have the same magnitude via Eq.~\ref{eq:mm_ll_bgo_corrections}. However, this equality does not generally hold. The lens breaks the isotropy of spacetime, potentially leading to the emergence of a preferred direction that could distinguish between the GW polarizations. In our particular case of interest, the geometric symmetry of the scattering setup we considered (particularly setting $z_s\rightarrow-\infty$ and $z_o\rightarrow+\infty$, as depicted in Fig~\ref{fig:propagation scheme}, left panel) combined with the additional assumptions discussed at the beginning of Sec.~\ref{Point like lens}, prevents any anisotropic effects from manifesting, extinguishing this potential distinction.
If observable, the polarization-dependence may be used to further characterize the lens-source configuration (i.e.~if the lens is near the source $|z_s|\ll |z_o|$).

Fixing the lens mass to $M_L=10 M_{\odot}$, the dependence of $\beta_{J}$ on its respective parameters is depicted in Fig.~\ref{fig:beta}. As shown in the aforementioned figure, $\beta_{J}$ is particularly sensitive to variations in the scalar charge.

The impact of these effects becomes larger for higher values of the scalar charge $q$ and for an impact parameter very close to the lens, typically within $2$ to Schwarzschild radii. In this regime, $\beta_{J}$ can reach high values, as can be seen from Fig.~\ref{fig:beta}, right panel. Further, it is worth mentioning that the $\beta_{mm}$ behavior for large values of the impact parameter scales as $\beta_{mm} \sim \tau R_s^2/b^3$ at the leading order; this result can be readily obtained by expanding Eq.~\eqref{eq:mm_ll_bgo_corrections} for large $b$ along with the $\beta_{J}$ definition of Eq.~\eqref{eq:beta_phase}.

Regardless of the chosen scalar charge value, these effects decrease rapidly with the impact parameter $b$. For a scalar charge of $q \sim 10^{-2}$ corresponding to $\tau=4.78\times 10^{-15}$ pc, (purple line, right panel, Fig.~\ref{fig:waveforms}), the scalar field behaves as a small perturbation to GR. For such a value, $\beta_{J}$ remains well below unity even if the scattering occurs at extremely close distances (just above one Schwarzschild radius).

The impact on the GW waveform is determined by $\beta_{J}/GMf$, resulting in a shift that is suppressed by several orders of magnitude in all the analyzed scenarios, making it difficult to distinguish deviations from GR.
Although tests of Brans-Dicke theory appear beyond the current capacity of GW detectors, we expect the signal to be enhanced by more complex mass distributions (beyond single point mass), where dispersive corrections from multiple objects may lead to a larger compound effects. 
The effect will also be enhanced by geometric time delays~\cite[Eq.~54]{Ezquiaga:2020dao}, not considered here.
Theories beyond Brans-Dicke will include new interactions, and are thus expected to increase the signal. These theories are also known to introduce additional scales beyond $\tau$: e.g.~the relevant scale for the scalar field in Kinetic Gravity Braiding theories~\cite{Deffayet:2007kf} is the Vainshtein radius $R_V$, which is many orders of magnitude larger than $\tau, R_S$~\cite{Ezquiaga:2020dao}. If dispersion is associated to $R_V$, its detection prospects will be strongly boosted. Ultimately, a full wave-optics framework will be necessary to address lenses with $f\sim GM_L$, for which dispersive corrections will be maximized.

\section{Conclusion}

\begin{table}[]
   \begin{tabularx}{\linewidth}{|Y|Y|Y|Y|Y|Y|Y|}
  \hline
   & $\delta\alpha^{(1)}_{nn}$ & $\delta\alpha^{(1)}_{mm}$ & $\delta\alpha^{(1)}_{ll}$ & $\delta\alpha^{(1)}_{ml}$ & $\delta\alpha^{(1)}_{nl}$ & $\delta\alpha^{(1)}_{nm}$ \\
  \hline
  GR & \textcolor{ForestGreen}{\Checkmark} & \textcolor{BrickRed}{\XSolidBrush} & \textcolor{BrickRed}{\XSolidBrush} & \textcolor{BrickRed}{\XSolidBrush} & \textcolor{BrickRed}{\XSolidBrush} & \textcolor{BrickRed}{\XSolidBrush} \\
  \hline
  BD & \textcolor{ForestGreen}{\Checkmark} & \textcolor{ForestGreen}{\Checkmark} & \textcolor{ForestGreen}{\Checkmark} & \textcolor{ForestGreen}{\Checkmark} & \textcolor{BrickRed}{\XSolidBrush} & \textcolor{BrickRed}{\XSolidBrush} \\
  \hline
\end{tabularx}
    \caption{\footnotesize{Schematic summary of the zero (\textcolor{BrickRed}{\XSolidBrush}) vs non-zero (\textcolor{ForestGreen}{\Checkmark}) bGO corrections in Einstein's GR and Brans-Dicke (BD), in the Jordan frame. In addition to introducing dispersive effects in the standard GW polarizations ($\delta\alpha_{ll}^{(1)},\delta\alpha_{mm}^{(1)}\neq 0$), BD modifies the $\delta\alpha_{nn}^{(1)}$. Moreover, the scalar field perturbation excites $\delta\alpha^{(1)}_{ml}$.}}
    \label{fig:table polarizations}
\end{table}

This work presents a novel framework to describe frequency-dependent corrections to the propagation of gravitational waves (GWs) in theories beyond Einstein's General Relativity (GR).
These represent dispersive phenomena, analog to how a prism splits light into its component frequencies. The role of the prism is played by a gravitational lens, which allows the standard GW polarizations $(+,\times)$ to interact with fields with lower spin ($e.g.,$ a scalar). Lens-induced dispersion (LID) phenomena appear beyond geometric optics (bGO) and can be tested on any GW signal, not requiring an electromagnetic counterpart.

We begin presenting a general, model-independent description of linearized GW propagation in inhomogeneous spacetime, relying only on a short-wave expansion.
For simplicity, we focus on scalar-tensor theories, which include an additional scalar field. 
The propagation of the radiative d.o.f. is described by a system of tensorial differential equations, coupled due to the interaction between gravity and the scalar field. We express this system compactly in matrix form and introduce the kinetic, amplitude and mixing matrices, which respectively encode interactions with two, one and zero derivatives of the perturbations, and whose coefficients are functions of the background quantities. We then describe how to define \textit{propagation eigenstates}, for which the kinetic matrix is diagonal~\cite{Ezquiaga:2020dao,Dalang:2020eaj}, and present the general structure of the equations in the short-wave expansion. The analogy here is similar to the way neutrinos change states as they travel. In the realm of beyond GR theories, the shift from interaction to propagation eigenstates involves interactions involving two derivatives, rather than zero.

We present the computation of bGO corrections in GR before addressing scalar-tensor theories (Sec.~\ref{GR}). This also serves as an introduction of the tetrad decomposition, affine parameter and luminosity distance. We review how bGO corrections introduce an apparent new polarization, the $nn$ mode, via dispersive corrections. We then apply the same concepts to Brans-Dicke (BD) theory (Sec.~\ref{BD}), presenting for the first time the structure of the equations that describe dispersive phenomena in GW propagation beyond GR. Although complex, two important properties of BD make these computations feasible: First, the diagonalization of the kinetic matrix is simple\til\eqref{BD diagonalization transformation}, and equivalent to the well-known definition of the Einstein frame metric.
Second, BD is a fully-luminal theory, with all excitations sharing the same propagation speed, wavevector and geodesics (Sec\til\ref{subsec:WKB-BD}).

We then present an explicit computation of lens-induced diffraction (LID) in BD in the presence of a point lens (Sec~\ref{Point like lens}). Additional, well-motivated approximations allow us to present analytical results: we assume the lens to be isolated and that deviations w.r.t. flat spacetime are small for the trajectories (Born approximation) and negligible at large separations from the lens.%
 As expected, all corrections reduce to GR results when the scalar field is constant, and LID vanish as $1/f$ at high frequencies, recovering the GO limit. To facilitate the interpretation of data in the context of BD theory, we rewrite our results in terms of the minimally-coupled Jordan frame metric, Eqs.\til\eqref{eq:GO-JF} and\til\eqref{eq:bGO-JF}.

Our results exemplify how LID provides a smoking gun for deviations from GR:
\begin{itemize}
    \item Additional physical polarizations are present and receive dispersive corrections, which depend on the theory parameters. For BD, the scalar field perturbation sources the $ml$ component, additional breathing mode which is not present in GR.%
    \item Like in GR, \textit{apparent} new polarizations ($nn$) are sourced by bGO corrections. The amplitude of this correction depends on the BD parameter.
    \item The standard metric polarizations ($+,\times$) receive novel frequency-dependent corrections. These are absent in GR and would be a clear signature of new gravitational dynamics.
    \item Frequency dependent corrections are common to $+,\times$ in our setting. Polarization-dependent dispersion occurs if the lens and source are near, and is expected generically in theories beyond BD.
\end{itemize}
This rich structure emerges even for the relatively simple case of BD, despite important simplifications like full-luminality and the absence of kinetic interactions. More complex theories of gravity will lead to more and/or stronger observational signatures.

Three important aspects of LID make them promising for testing gravity. First, the frequency-dependent corrections can be tested on all GW sources (irrespective of their intrinsic properties, electromagnetic counterpart, etc.).
Second, they stem from inhomogeneous backgrounds, which distinguish between the standard GW polarizations ($+,\times$).
Finally, they arise from interactions with zero derivatives, which are expected in any modified theory. LID phenomena are therefore a universal prediction in theories beyond GR.

This work 
bridges important gaps in the theoretical understanding of GWs propagation beyond GR. Future research 
should address more
complex gravity theories, %
including different speeds for scalar and tensor waves, screening mechanisms, or non-trivial cosmological dynamics,
likely leading to enhanced dispersive effects on the gravitational wave signal.
Because dispersive effects are stronger at low frequencies, this program needs to not only consider ground detectors, but also space-borne observatories and ultra-low frequency GWs observable through pulsar-timing arrays.
This program will enable novel tests of gravity and dark energy theories, leveraging the full potential of the GW spectrum.

\subsection*{Acknowledgments}
We thank Guillerme Brando, Han Gil Choi, Giulia Cusin, Charles Dalang, Jose Maria Ezquiaga, Srashti Goyal, Serena Giardino, Stefano Savastano, and Hiroki Takeda for the discussions and comments on this work. N.M. thanks D. Usseglio for his valuable insights into the project and is especially grateful to Prof. S. Capozziello for his support and encouragement throughout all stages of this work. N.M. also acknowledges the Scuola Superiore Meridionale for sponsoring his visiting period to the Albert Einstein Institute (AEI) in Potsdam, which provided an invaluable opportunity for collaboration and research.
\appendix
\section{Weak field limit}\label{appendix: Weak field limit}
Given the line element of Eq.\til\eqref{wfl metric}, one gets the Christoffel symbols up to first order in $\Psi$

\begin{align}
    \delta\tensor{\Gamma}{^{0}_{00}}&=\delta\tensor{\Gamma}{^{0}_{ij}}=\delta\tensor{\Gamma}{^{i}_{j0}}=0,\\
    \delta\tensor{\Gamma}{^{0}_{i0}}&=\partial_{i}\Psi,\\
    \delta\tensor{\Gamma}{^{i}_{jk}}&=\delta_{jk}\partial^i\Psi-\delta^i_k\partial_j\Psi-\delta^i_j\partial_k\Psi,
\end{align}
where $\delta_{ij}$ is the Kronecker delta. Further, the linearized Riemann and Ricci tensors and scalar, respectively, are then provided
\begin{align}
    \delta \tensor{R}{^\rho _{\mu\lambda\nu}}&=\partial_{\lambda}\delta\tensor{\Gamma}{^{\rho}_{\mu\nu}}-\partial_{\nu}\delta\tensor{\Gamma}{^{\rho}_{\mu\lambda}},\\
    \delta R_{\mu\nu}&=\delta^{\lambda}_{\rho}\delta \tensor{R}{^\rho _{\mu\lambda\nu}},\\
    \delta R&=g^{\mu\nu}\delta \tensor{R}{_{\mu\nu}}.
\end{align}

\section{Brans-Dicke background functions}\label{Terms Brans Dicke}
Here we present the non-zero components of the kinetic, amplitude, and mass matrices, appearing in Eq.\til\eqref{eq: full progation Brans-Dicke}, which constitute the system of differential equations describing the propagation of GWs and scalar waves in the TT gauge. In particular, we have

\begin{align}
\label{BD effective metric tensor}
\tensor{\mathsf{\Tilde{K}}}{_\mu_\nu^\alpha^\beta^\gamma^\rho}&\equiv-\frac{1}{2}\bar{\phi}\delta^{\alpha}_{\mu}\delta^{\beta}_{\nu}g^{\gamma\rho},\\ 
\label{eq: BD effective metric}
\mathsf{\tilde{K}}^{\gamma\rho}&\equiv-\Bar{\phi}g^{\gamma\rho},\\ 
\label{eq: A BD}
\tensor{\mathsf{\Tilde{A}}}{_\mu_\nu^\alpha^\beta^\gamma}&\equiv\frac{1}{2}\left(2\Bar{\phi}^{\alpha}\delta^{\beta}_{(\mu}\delta^{\gamma}_{\nu)}-\Bar{\phi}^{\gamma}\delta^{\alpha}_{\mu}\delta^{\beta}_{\nu}\right),\\ \label{eq: Q BD}
\tensor{\mathsf{\Tilde{A}}}{_\mu_\nu^\gamma}&\equiv\frac{3+2\omega}{2\Bar{\phi}}\left(g_{\mu\nu}\Bar{\phi}^{\gamma}-2\Bar{\phi}_{(\mu}\delta^{\gamma}_{\nu)}\right),\\ \label{A gamma}
\tensor{\mathsf{\Tilde{A}}}{^\gamma}&\equiv\Bar{\phi}^{\gamma},\\
\begin{split}
    \tensor{\mathsf{\Tilde{M}}}{_{\mu\nu}^{\alpha\beta}}&\equiv\frac{\Bar{\phi}}{2}\left(g_{\mu\nu}R^{\alpha\beta}+2\tensor{R}{_{(\mu}^\alpha}\delta^{^\beta}_{\nu)}-\delta^{\alpha}_{\mu}\delta^{\beta}_{\nu}R+2\tensor{R}{^\alpha_{\mu\nu}}^{\beta}\right)+
    \\&+\delta^{\alpha}_{\mu}\delta^{\beta}_{\nu}\left(\Box\Bar{\phi}-\frac{\omega X}{\Bar{\phi}}\right)-g_{\mu\nu}\left(\Bar{\phi}^{\beta\alpha}+\frac{1}{2}\Bar{\phi}^{\alpha}\Bar{\phi}^{\beta}\right),
\end{split}
\\
\begin{split}
  \mathsf{\Tilde{M}}_{\mu\nu}^{h}&\equiv R_{\mu\nu}-g_{\mu\nu}\left[\frac{1}{2}R-\frac{\Box\Bar{\phi}}{\Bar{\phi}}-\frac{(3+\omega)X}{\Bar{\phi}^2}\right]+\\&+\frac{(3+\omega)}{\Bar{\phi}^2}\Bar{\phi}_{\mu}\Bar{\phi}_{\nu}-\frac{\Bar{\phi}_{\nu\mu}}{\Bar{\phi}},  
\end{split}
\\ 
\mathsf{\Tilde{M}}^{\alpha\beta}_{\phi}&\equiv\Bar{\phi}(\Bar{\phi}_{\alpha\beta}),\\
\mathsf{\Tilde{M}}&\equiv\frac{2X}{\Bar{\phi}},
\end{align}
where $X\equiv-\nabla_{\mu}\bar{\phi}\nabla^{\mu}\bar{\phi}/2$ is the canonical kinetic term of the scalar field.

\section{Functions for bGO corrections}\label{Function of the modes}
Here, we present the functions to be integrated, along the $z$-axis, to obtain the bGO corrections. For clarity, we have set $\Bar{\phi}_{\infty}=1$, as discussed in Sec.\til\ref{PLL Brans Dicke}. Let us start by showing the functions of Eq.\til\eqref{scalar amplitude correction}
\begin{equation}
    f^{\phi}(b,z_s,\tau;z)=\frac{\bar{D}(z_s)}{\sqrt{\Bar{\phi}(z_s)}}\frac{p(b,z_s,\tau;z)}{q(b,z_s,\tau;z)},
\end{equation}
with
\begin{equation}
\begin{split}
        p&(b,z_s,\tau;z)\equiv \tau^2 \left(8 b^2+9 z^2-2 z z_s+z_s^2\right)+\\&+16 \tau (b^2+z^2)^{3/2}+8 \left(b^2+z^2\right)^2,
\end{split}
\end{equation}
\begin{equation}
\begin{split}
q&(b,z_s,\tau;z)]\equiv4\left(b^2+z^2\right) (z-z_s)^2 \times\\&\times\left[\left(b^2+z^2\right)+2 \tau (b^2+z^2)^{1/2} +\tau^2\right].
\end{split}
\end{equation}
The functions multiplying $\tilde{\zeta}^{(0)}_{+}$ and $\tilde{\zeta}^{(0)}_{-}$, respectively, are
\begin{align}\label{scalarplus}
    f^{+}(b,z_s,\tau;z)&=-\bar{D}(z_s)\sqrt{\Bar{\phi}(z_s)}\frac{3 \tau (b^2+2zR_s) }{2g(b,\tau;z)},\\ \label{scalarminus}
    f^{-}(b,z_s,\tau;z)&=0,
\end{align}
with
\begin{equation}
\begin{split}
g(b,\tau;z)\equiv\left(b^2+z^2\right)^2 \left[\tau+(b^2+z^2)^{1/2}\right].
    \end{split}
\end{equation}

We proceed by presenting the elements that compose Eq.\til\eqref{perturbed tensor in z}, only considering the coefficients appearing in Eq.\til\eqref{perturbed bgo tensor amplitude}. The quantity $\delta\alpha^{(1)}_{nn}$ is build up by 
\begin{equation}
\begin{split}
   f^{\phi}_{nn}&(b,\tau;z)=2\Omega^2R_s\frac{\bar{D}(z_s)}{\sqrt{\Bar{\phi}(z_s)}}\times\\&\times\left[\frac{  b^2 \tau}{(b^2+z^2)^{1/2}g(b,\tau;z)}\right.\left.-\frac{\left(b^2-2 z^2\right)}{\left(b^2+z^2\right)^{5/2}}\right],
\end{split}
\end{equation}
\begin{equation}
\begin{split}
    f^{+}_{nn}&(b,\tau;z)=-b^2\Omega^2R_s\bar{D}(z_s)\sqrt{\Bar{\phi}(z_s)}\times\\&\times\left[\frac{ \tau  }{(b^2+z^2)^{1/2}g(b,\tau;z)}\right.\left.+\frac{3  }{\left(b^2+z^2\right)^{5/2}}\right],
\end{split}
\end{equation}
\begin{equation}
    f^{-}_{nn}(b,\tau;z)=0.
\end{equation}
Proceeding with $\delta\alpha^{(1)}_{mm}$, we have

\begin{equation}
\begin{split}
   f^{\phi}_{mm}&(b,z_s,\tau;z)=b^2R_s\frac{\bar{D}(z_s)}{\sqrt{\Bar{\phi}(z_s)}}\times\\&
   \times\left[-\frac{ \tau }{(b^2+z^2)^{1/2}g(b,\tau;z)}+\frac{3 }{\left(b^2+z^2\right)^{5/2}}\right],
\end{split}
\end{equation}
\begin{equation}
\begin{split}
  f^{+}_{mm}&(b,z_s,\tau;z)=f^{-}_{mm}(b,z_s,\tau;z)=\\&=R_s\bar{D}(z_s)\sqrt{\Bar{\phi}(z_s)}\left\{\frac{plo\left(b^2-2 z^2\right) }{2  \left(b^2+z^2\right)^{5/2}}-\frac{\tau(b^2+z^2)^{1/2}}{g(b,\tau;z)}+\right.\\&+\left.\frac{z\tau^2\left(b^2+z^2\right)^2}{4g(b,\tau;z)^2}+\frac{b^2\tau}{2(b^2+z^2)^{1/2}g(b,\tau;z)}\right\},
 \end{split}
\end{equation}
Given that the tetrads $m^{\mu}$ and $l^{\mu}$ differ only by a complex conjugation, the functions characterizing $f^{\phi,\pm}_{ll}(b,z_s,\tau;z)$ exhibit the same functional dependence as those of $\delta\alpha^{(1)}_{mm}$,
following
\begin{align}
    \begin{split}
    f^{\phi}_{mm}(b,z_s,\tau;z)=&f^{\phi}_{ll}(b,z_s,\tau;z),\\
    f^{+}_{mm}(b,z_s,\tau;z)=&f^{+}_{ll}(b,z_s,\tau;z),\\ f^{-}_{mm}(b,z_s,\tau;z)=&-f^{-}_{ll}(b,z_s,\tau;z).
    \end{split}
\end{align}
The functions for $\delta\alpha^{(1)}_{nm}$ are instead
\begin{equation}\label{nm function scalar}
\begin{split}
       f^{\phi}_{nm}&(b,z_s,\tau;z)=\sqrt{2}\Omega b z R_s\frac{\bar{D}(z_s)}{\sqrt{\Bar{\phi}(z_s)}}\times\\&\times\left[\frac{\tau}{(b^2+z^2)^{1/2}g(b,\tau;z)}-\frac{3}{(b^2+z^2)^{5/2}}\right].
\end{split}
\end{equation}
\begin{equation}\label{nm function tensor}
    \begin{split}
        f^{+}_{nm}&(b,z_s,\tau;z)=f^{-}_{nm}(b,z_s,\tau;z)=\\&-\frac{\Omega   R_sb}{2\sqrt{2}}\bar{D}(z_s)\sqrt{\Bar{\phi}(z_s)}
       \left[\frac{3z}{(b^2+z^2)^{5/2}}+\right.\\&\left.+\frac{1}{(b^2+z^2)^{3/2}(z-z_s)}+\frac{z\tau}{(b^2+z^2)^{1/2}g(b,\tau;z)}\right],
    \end{split}
\end{equation}
Analogously, the functions $f^{\phi,\pm}_{nm}(b,z_s,\tau;z)$ follows
\begin{align}
    \begin{split}
    f^{\phi}_{nl}(b,z_s,\tau;z)=&f^{\phi}_{nl}(b,z_s,\tau;z),\\
    f^{+}_{nl}(b,z_s,\tau;z)=&f^{+}_{ll}(b,z_s,\tau;z),\\ f^{-}_{nl}(b,z_s,\tau;z)=&-f^{-}_{nl}(b,z_s,\tau;z).
    \end{split}
\end{align}

We now consider the case where $z_s\rightarrow-\infty$ and $z_o\rightarrow+\infty$: apart from the integration boundaries, this operation also applies to the functions $f^{\phi,\pm}(b,z_s,\tau;b)$ and $f_{AB}^{\phi,\pm}(b,z_s,\tau;b)$. In particular, we observe that the functional dependence on $z_s$ appears only in the functions $f^{\phi}(b,z_s,\tau;b)$ and $f^{\pm}_{nm}(b,z_s,\tau;b)$. Performing the aforementioned limit on these two, we obtain
\begin{align}\label{scalar phi limit zs}
\begin{split}
       f^{\phi}_{\infty}&(b,\tau;z)=\frac{\bar{D}(z_s)}{\sqrt{\Bar{\phi}(z_s)}}\frac{\tau^2(b^2+z^2)}{4g(b,\tau;z)^2},
\end{split}
\end{align}
\begin{align}
\begin{split}
       f^{\pm}_{nm_{\infty}}(b,\tau;z)=&-\frac{\Omega   R_sb}{2\sqrt{2}}\bar{D}(z_s)\sqrt{\Bar{\phi}(z_s)}\times
       \\&\times\left[\frac{3z}{(b^2+z^2)^{5/2}}\right.\left.+\frac{z\tau}{(b^2+z^2)^{1/2}g(b,\tau;z)}\right].
\end{split}
\end{align}

\begin{figure}[h!]
\centering
{\includegraphics[width=\linewidth]{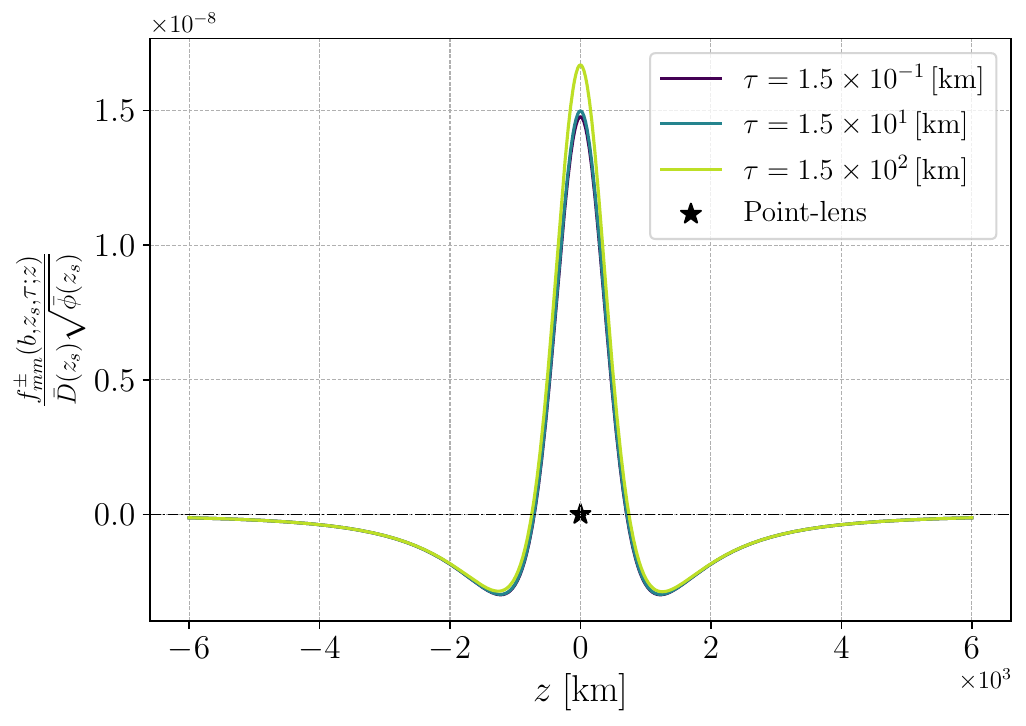}\label{fig:subfig2}}\hfill
\caption{Plot of the integrand function $f^{\pm}_{mm}(b,zs, \tau, z)$ as a function of $z$, with $b$ fixed at $10^3$ km for different values of $\tau\equiv q G M_L$, corresponding to varying scalar charge values $q$. $M_L=10M_{\odot}$ The black star at $z = 0$ represents the location of the point-lens.}
  \label{fig:fmm_integrando}
\end{figure}

\subsection{Results from analytic integration}\label{Analytic integrals}
We define $F^{\phi,\pm}(b,\tau)$ and $F^{\phi,\pm}_{AB}(b,\tau)$ as
\begin{align}\label{scalar integrals app}
    F^{\phi,\pm}(b,\tau)&\equiv\int_{-\infty}^{+\infty}\frac{dz}{\Omega}f^{\phi,\pm}_{\infty}(b,\tau;z),\\ \label{tensor integrals app}
    F^{\phi,\pm}_{AB}(b,\tau)&\equiv\int_{-\infty}^{+\infty}\frac{dz}{\Omega}f^{\phi,\pm}_{AB_{\infty}}(b,\tau;z).
\end{align}
The scalar bGO correction $\delta\phi^{(1)}$ is obtained by employing the definition\til\eqref{scalar integrals app} along with Eqs.\til\eqref{scalarplus},\til\eqref{scalarminus} and\til\eqref{scalar phi limit zs}, thus obtaining
\begin{equation}
\begin{split}
        F^{\phi}(b,\tau)&\equiv\frac{\bar{D}(z_s)}{\sqrt{\Bar{\phi}(z_s)}}\left[\frac{\pi b^2-2b \tau-\pi \tau^2}{4b(b^2-\tau^2)}+\frac{b^2-2\tau^2}{(b^2-\tau^2)}u(b,\tau)\right],
\end{split}
\end{equation}
\begin{equation}
    \begin{split}
        F^{+}(b,\tau)\equiv&-\bar{D}(z_s)\sqrt{\Bar{\phi}(z_s)}\times\\&\times\frac{3b^2}{4\tau^2}\left[\frac{2\pi b^2-4b \tau+\pi \tau^2}{b^3}+8u(b,\tau)\right],
    \end{split}
\end{equation}

\begin{equation}
    \begin{split}
        F^{-}(b,\tau)\equiv0,
    \end{split}
\end{equation}
with 
\begin{equation}
    \begin{split}
        u(b,\tau)&\equiv\frac{1}{(b^2-\tau^2)^{1/2}}\times\\&\times\left[\tan^{-1}\left(\frac{\tau}{\sqrt{b^2-\tau^2}}\right)-\cot^{-1}\left(\sqrt{\frac{b-\tau}{b+\tau}}\right)\right].
    \end{split}
\end{equation}
The limit to GR is achieved by setting $\tau \rightarrow 0$, which leads $F^{\phi}(b,\tau)$ and $F^{+}(b,\tau)$ to  identically vanish. This result is consistent since, in GR, the scalar wave does not exist.

The results regarding the coefficients $\delta\tilde{\alpha}^{(1)}_{AB}$ composing the tensor bGO correction $h^{(1)}_{\mu\nu}$ are derived by using Eq.\til\eqref{tensor integrals app}.
By starting from the $\delta\tilde{\alpha}^{(1)}_{nn}$, the result is 
\begin{equation}
\begin{split}
        F^{\phi}_{nn}(b,\tau)&=\frac{\bar{D}(z_s)}{\sqrt{\Bar{\phi}(z_s)}}\frac{\Omega R_s}{\tau^3}\times\\&\times\left\{\frac{8\tau^3}{3b^2}-\frac{\pi\tau^2}{b}+4\tau-2b\left[\pi+4bu(b,\tau)\right]\right\},
\end{split}
\end{equation}
\begin{equation}
    \begin{split}
         F^{+}_{nn}(b,\tau)=&-\bar{D}(z_s)\sqrt{\Bar{\phi}(z_s)}\frac{4\Omega R_s}{b^2}\times\\&\times\left\{\frac{4}{3}-\frac{\pi b   }{8\tau}+\frac{b^2}{2\tau^2}-\frac{\pi b^3}{4\tau^3}-\frac{b^4}{\tau^3}u(b,\tau)\right\},
    \end{split}
\end{equation}
\begin{equation}
    \begin{split}
         F^{-}_{nn}&(b,\tau)=0.
    \end{split}
\end{equation}
The $\tau\rightarrow0$ limit makes $F^{\phi}_{nn}(b,\tau)$ and $F^{-}_{nn}(b,\tau)$ vanish but not $F^{+}_{nn}(b,\tau)$ 
whose terms in the parenthesis resemble to provide the result in Eq.\til\eqref{GR nn mode}, with $\bar{\phi}(z_s)\rightarrow1$. 

The result for the $mm$ and $ll$ components acts as
\begin{equation}
    \begin{split}
         F^{\phi}_{mm}(b,\tau)=&F^{\phi}_{ll}(b,\tau)=\frac{\bar{D}(z_s)}{\sqrt{\Bar{\phi}(z_s)}}\frac{R_s}{\Omega}\times\\&\times\left[\frac{8}{3b^2}+\frac{\pi}{2b\tau}-\frac{2}{\tau^2}+\frac{\pi b}{\tau^3}+\frac{4b^2u(b,\tau)}{\tau^3}\right],
    \end{split}
\end{equation}

\begin{align}\label{eq:mm_ll_bgo_corrections}
    \begin{split}
         &F^{+}_{mm}(b,\tau)=F^{-}_{mm}(b,\tau)=F^{+}_{ll}(b,\tau)=-F^{-}_{ll}(b,\tau)=\\&=\bar{D}(z_s)\sqrt{\Bar{\phi}(z_s)}\frac{R_s}{\Omega}\times\\&\times\left[-\frac{4}{3b^2}+\frac{3\pi}{4b\tau}+\frac{1}{\tau^2}-\frac{\pi b}{2\tau^3}-\frac{2(b^2-2\tau^2)u(b,\tau)}{\tau^3}\right].
    \end{split}
\end{align}
To conclude, the integration of Eqs.\til\eqref{nm function scalar} and\til\eqref{nm function tensor} vanishes identically, thus providing $\delta\tilde{\alpha}^{(1)}_{nm}=\delta\tilde{\alpha}^{(1)}_{nl}=0$.

\bibliography{main.bbl}
\end{document}